\newif\ifarxiv

\arxivtrue

\ifarxiv

\documentclass[english,reprint,nofootinbib,notitlepage]{revtex4-1}

\else
\fi

\usepackage{newtxtext}

\usepackage[utf8]{inputenc}
\usepackage{verbatim}
\usepackage{amsmath}
\usepackage{amsthm}
\usepackage{amssymb}
\usepackage{stmaryrd}
\usepackage{graphicx}
\usepackage{xargs}[2008/03/08]
\usepackage{microtype}
\usepackage[hidelinks]{hyperref}
\usepackage{xspace}
\usepackage{color}  %
\RequirePackage[dvipsnames,usenames]{xcolor}
\usepackage[capitalize]{cleveref}
\usepackage{tikz}
\usetikzlibrary{shapes.misc,intersections,arrows}

 \usepackage[flushmargin,bottom]{footmisc}

\ifarxiv
\makeatletter
\fi

\DeclareFontFamily{U}{mathc}{}
\DeclareFontShape{U}{mathc}{m}{it}%
{<->s*[1.03] mathc10}{}
\DeclareMathAlphabet{\mathcal}{U}{mathc}{m}{it}

\usepackage{enumitem}
\setlist[enumerate]{leftmargin=15pt,topsep=3pt,itemsep=0pt,partopsep=3pt,parsep=3pt}

\newcommand{\ft}{\tau}
\global\long\def\sysX{$\mathcal{X}$\xspace}
\global\long\def\sysY{$\mathcal{Y}$\xspace}

\global\long\def\KL{D_{\text{KL}}}
\global\long\def\argmax{\operatornamewithlimits{argmax}}
\global\long\def\argmin{\operatornamewithlimits{argmin}}
\global\long\def\p{p}

\newcommandx\TE[2][usedefault, addprefix=\global, 1=p, 2=t]{\mathcal{T}_{#1}(Y_{#2}\shortrightarrow X_{#2+1})}
\global\long\def\ft{\tau}

\global\long\def\cg{\phi}
\global\long\def\Cg{\Phi}
\global\long\def\Aset{\mathcal{A}}
\global\long\def\pft{\p_{X_\ft}}
\global\long\def\cfbase{\hat{p}}

\global\long\def\pcg{\cfbase^{\cg}}
\global\long\def\pcgft{\pcg_{X_\ft}}
\global\long\def\pfft{\pfull_{X_{\ft}}}

\newcommandx\MI[2][usedefault, addprefix=\global, 1=p, 2=]{I_{#1}(X_{#2},Y_{#2})}

\global\long\def\pfull{\cfbase^{\text{full}}}
\global\long\def\ktlntwo{k_{B}T\ln2}

\global\long\def\pcs{\cfbase^{\text{opt}}}

\global\long\def\SS{\mathcal{S}_{\text{stored}}}
\global\long\def\sS{\mathcal{s}_{\text{stored}}}
\global\long\def\SO{\mathcal{S}_{\text{observed}}}
\global\long\def\sO{\mathcal{s}_{\text{observed}}}
\global\long\def\DS{\mathcal{D}_{\text{stored}}}
\global\long\def\DO{\mathcal{D}_{\text{observed}}}
\newcommandx\Vtot[1][usedefault, addprefix=\global, 1=]{\Delta V_{\text{tot}}^{#1}}
\global\long\def\VtotO{\Vtot[\text{observed}]}
\global\long\def\VtotS{\Vtot[\text{stored}]}
\global\long\def\nS{\kappa_{\text{stored}}}

\global\long\def\pS{\eta_{\text{stored}}}
\global\long\def\pO{\eta_{\text{observed}}}

\begin{document}

\title{Semantic information, autonomous agency, and nonequilibrium statistical physics}

\ifarxiv

\author{Artemy Kolchinsky\\
{\small \url{artemyk@gmail.com}}}
\affiliation{Santa Fe Institute, Santa Fe, NM 87501, USA}
\author{David H. Wolpert\\
{\small \url{david.h.wolpert@gmail.com}\\
 \url{http://davidwolpert.weebly.com}}}
\affiliation{Santa Fe Institute, Santa Fe, NM 87501, USA\\
Massachusetts Institute of Technology\\
Arizona State University}

\else
\author{
Artemy Kolchinsky$^{1}$ and David H. Wolpert$^{1,2,3,4}$}

\address{$^{1}$Santa Fe Institute, Santa Fe, NM 87501, USA\\
$^{2}$ Massachusetts Institute of Technology\\
$^{3}$ Arizona State University\\
$^{4}$ {\small \url{http://davidwolpert.weebly.com}}}

\subject{xxxxx, xxxxx, xxxx}

\keywords{xxxx, xxxx, xxxx}

\corres{Artemy Kolchinsky\\
\email{artemyk@gmail.com}}

\fi

\begin{abstract}
Shannon information theory provides various measures of so-called ``syntactic information'', which reflect the amount of statistical correlation between systems. In contrast, the concept of ``semantic information'' refers  to those correlations which carry significance or ``meaning'' for a given system. Semantic information plays an important role in many fields, including biology, cognitive science, and philosophy, and there has been a long-standing interest in formulating a broadly applicable and formal theory of semantic information. In this paper we introduce such a theory. We define semantic information as the syntactic information that a physical system has about its environment which is causally necessary for the system to maintain its own existence.  ``Causal necessity'' is defined in terms of counter-factual interventions which scramble correlations between the system and its environment, while ``maintaining existence'' is defined in terms of the system's ability to keep itself in a low entropy state. We also use recent results in nonequilibrium statistical physics to analyze semantic information from a thermodynamic point of view. Our framework is grounded in the intrinsic dynamics of a system coupled to an environment, and is applicable to any physical system, living or otherwise.  It leads to formal definitions of several concepts that have been intuitively understood to be related to semantic information, including ``value of information'', ``semantic content'', and ``agency''.
\end{abstract}

\maketitle

\section{Introduction\label{sec:Introduction}}

The concept of \emph{semantic information} refers to information which is in some sense meaningful for a system, rather than merely correlational.
It
plays an important role in many fields, including biology~\cite{Dretske1981-oc,shea_representation_2007,godfrey-smith_biological_2016,godfrey-smith_information_2007,collier_information_2008,sterelny_sex_1999,millikan1984language,jablonka_information:_2002,weinberger_theory_2002}, cognitive
science~\cite{barham_dynamical_1996,ruiz-mirazo_basic_2004,moreno_agency_2005,deacon_2008,bickhard_autonomy_2000}, artificial intelligence~\cite{harnad1990symbol,glenberg2000symbol,steels2003evolving}, information theory~\cite{crutchfield_semantics_1992,tishby2000information,polani_relevant_2006,vitanyi2006meaningful}, and philosophy~\cite{Bar-Hillel1953-ph,Dennett1983-dg,sep-information-semantic}\footnote{Semantic information has also sometimes been called  ``meaningful information''
\cite{nehaniv_meaning_1999,nehaniv_meaningful_2003,atlan1987self,reading2011meaningful}, ``relevant
information''~\cite{tishby2000information,polani_relevant_2006},
``functional information''~\cite{farnsworth_living_2013,sharov_functional_2010}, and ``pragmatic information''~\cite{weinberger_theory_2002} in the literature.}. Given the ubiquity of this concept, an important question is whether it
can be defined in a formal and broadly
applicable manner. Such a definition could be used to %
 analyze and clarify issues concerning semantic information in a variety of fields, and possibly to uncover
novel connections between those fields.
A second, related question is whether one can construct a formal
definition of semantic information that  applies not only to living beings,
but \emph{any} physical
system --- whether a rock, a hurricane, or a cell. A formal
definition which can be applied to the 
full range of physical systems may
provide novel insights into how living and nonliving systems are related.

The main contribution of this paper is a definition of semantic information that 
positively answers both
of these questions,
following ideas publicly presented at the FQXi's 5th International Conference~\cite{fqxitalk_2016} and explored by Carlo Rovelli~\cite{rovelli_meaning_2018}. 
In a nutshell, we define 
\emph{semantic
information} as \emph{the information that a physical system has about its environment
that is causally necessary for the system to maintain its own existence over time}.
Our definition is grounded in the intrinsic dynamics of a system and its environment, and, as we will show, it 
formalizes existing intuitions while leveraging ideas from information theory and nonequilibrium statistical physics~\cite{seifert2012stochastic,parrondo2015thermodynamics}.   It also leads to a non-negative decomposition of information measures into ``meaningful bits'' and ``meaningless bits'', and provides a coherent quantitative framework for expressing a constellation
of concepts related to ``semantic information'', such as ``value
of information'', ``semantic content'', and ``agency''.  

\subsection{Background}
\label{subsec:background}

Historically, {semantic information} 
has been contrasted with \emph{syntactic information}, which quantifies
various kinds of statistical correlation between two systems, with
no consideration of what such correlations ``mean''. 
Syntactic
information is usually studied using Shannon's well-known information
theory and its extensions~\cite{shannon1948mathematical,cover_elements_2012},
 which provide measures that quantify how much knowledge of the
state of one system reduces statistical uncertainty about the state
of the other system, possibly at a different point in time. When 
introducing his information theory, Shannon focused on the engineering problem of accurately transmitting
messages across a telecommunication channel, and explicitly sidestepped 
 questions regarding what meaning, if any, the messages might have~\cite{shannon1948mathematical}.

How should we fill in the gap that Shannon explicitly introduced?
One kind of approach ---
common in economics, game theory, and statistics --- begins
by assuming an idealized system that pursues some externally-assigned
goal, usually formulated as the optimization of an objective function, such as  utility~\cite{vomo44,futi91,polani_information-theoretic_2001,gould_risk_1974,hess_risk_1982},
distortion~\cite{cover_elements_2012}, or prediction error~\cite{tishby2000information,harremoes_information_2007,degroot_uncertainty_1962,jiao_justification_2015}.
Semantic information is then defined as information which helps the
system to achieve its goal (e.g., information about tomorrow's stock
market prices would help a trader increase
their economic utility). Such approaches can be quite useful and have
lent themselves to important formal developments. However, they have
the major shortcoming that they specify the goal of the system \emph{exogenously}, meaning that they are not appropriate for grounding meaning in the \emph{intrinsic} properties
of a particular physical system. The semantic information they quantify
has meaning {for the external scientist} who imputes goals to the system, rather than for the system
itself. 

In biology, the goal of an organism is often considered to be evolutionary success (i.e., the maximization of fitness), which has led to the so-called \emph{teleosemantic}  approach to semantic information. %
Loosely speaking, teleosemantics proposes that a biological trait carries semantic information if the presence of the trait was ``selected for'' because, in the evolutionary past, the trait correlated with particular states of the environment~\cite{Dretske1981-oc,shea_representation_2007,godfrey-smith_biological_2016,godfrey-smith_information_2007,collier_information_2008,sterelny_sex_1999,millikan1984language}.
To use a well-known
example, when a frog sees a small black spot in its visual field,
it snaps out its tongue and attempts to catch a fly. 
This stimulus-response behavior was selected for, since small black spots in the visual field correlated
with the presence of flies and eating flies was good for frog fitness.
Thus, a small black spot in the visual field of a frog has semantic
information, and refers to the presence of flies. 

While in-depth discussion
of teleosemantics is beyond the scope of this paper, we note that 
some of its central features make it deficient for our purposes.
First, it is only applicable to physical systems
that undergo natural selection. Thus, it is not clear how to apply it to entities like non-living systems, protocells, or synthetically-designed organisms. %
Moreover, teleosemantics is ``etiological''~\cite{schlosser_self-re-production_1998,mossio_organizational_2009}, meaning that it defines semantic information in terms of the {past} history
of a system. Our goal is to develop a theory of semantic information that is based purely in the intrinsic dynamics of a system in a given environment, irrespective of the system's origin and past history.

Finally, another approach to semantic information comes from literature on so-called
\emph{autonomous agents}~\cite{maturana1991autopoiesis,schlosser_self-re-production_1998,bickhard_autonomy_2000,ruiz-mirazo_basic_2004,moreno_agency_2005,Thompson2008-da,Froese2009-wz,mossio_organizational_2009}.
An autonomous agent is a far-from-equilibrium system which actively maintains its own existence within some environment~\cite{raymond_communication_1950,collier_intrinsic_1990,collier_complexly_1999,bickhard_autonomy_2000,ruiz-mirazo_basic_2004,moreno_agency_2005,deacon_2008,friston_free-energy_2007,nehaniv_meaning_1999,kauffman_molecular_2003}. A prototypical example of an autonomous
agent is an organism, but in principle the notion can also be
applied to robots~\cite{maes1990designing,melhuish2006energetically}
and other
non-living systems \cite{mcgregor2009life,barandiaran2009defining}. For an autonomous
agent, self-maintenance is a fundamentally intrinsic goal, which is neither assigned by an external scientist analyzing the system, nor based on past evolutionary
history. 

In order to maintain themselves, autonomous agents must typically {observe} (i.e., acquire information about) their environment, and then respond in different and ``appropriate'' ways. For instance, a chemotactic bacterium
senses the direction of chemical gradients in its particular
environment and then moves in the direction of those gradients, thereby locating food and maintaining its own existence.  In this sense, autonomous agents can be distinguished from ``passive''
self-maintaining structures that emerge whenever appropriate boundary
conditions are provided, such as Bénard cells~\cite{chandrasekhar1961hydrodynamic} and some other well-known nonequilibrium systems.

Research on autonomous agents suggests that information about the environment that is used by an autonomous agent for self-maintenance is intrinsically meaningful~\cite{barham_dynamical_1996,bickhard_autonomy_2000,ruiz-mirazo_basic_2004,moreno_agency_2005,Thompson2007-qm,Thompson2008-da,deacon_2008,Froese2009-wz,nehaniv_meaning_1999,nehaniv_meaningful_2003}. 
However, until now, such ideas 
have remained largely informal. %
In particular, there has been
no formal proposal in the autonomous agents literature for quantifying the amount of semantic information possessed by
{any} given physical system, nor for identifying the meaning (i.e., the semantic content) of particular system states.

\subsection{Our contribution}

We propose a formal, intrinsic definition of semantic
information, applicable to any physical system coupled to an external environment, whether a rock, a hurricane, a bacterium, or a sample from an alien
planet.\footnote{Much of our 
approach can also be used to quantify semantic information
in any dynamical system, not just physical systems. For the purposes
of this paper, however, we focus our attention on physical systems.}
%

We assume the following setup:
there is a physical world which can be decomposed into two subsystems, %
which we refer to as ``the system \sysX'' and ``the environment \sysY''
respectively. 
We suppose that at some initial time $t=0$, the system and environment are jointly distributed according to some initial distribution $\p(x_{0},y_{0})$. They then undergo coupled (possibly stochastic) dynamics until time $\ft$, 
where $\ft$ is some timescale of interest. 

Our goal is to define the semantic information that the
system has about the environment.  To do so, we make use of 
a \textbf{viability function}, a  real-valued function which quantifies the system's ``degree
of existence'' at a given time. 
While there are several possible ways to define a viability function,
in this paper we take inspiration from statistical physics
\cite{touchette_information-theoretic_2000,touchette2004information,cao_thermodynamics_2009}
and define the viability function as the negative Shannon entropy
of the distribution over the states of system \sysX. This choice is motivated by the fact that Shannon entropy provides an upper bound on the probability that the system occupies any small set of ``viable'' states~\cite{ashby_design_1960,beer_dynamical_1996,di_paolo_autopoiesis_2005,agmon_exploring_2016}. We are also motivated by the connection between Shannon entropy and thermodynamics~\cite{parrondo2015thermodynamics,esposito2011second,seifert2012stochastic,maroney2009generalizing,wolpert_landauer_2016a,horowitz_minimum_2017}, which allows us to connect our framework to results in nonequilibrium statistical physics.
Further discussion of this viability function, as well as other possible viability functions, is found in \cref{subsec:The-viability-function}.

%
%
%
%
%
%
%
%
%

Information theory provides many measures of
the syntactic information shared between the system and its environment. For any particular measure of syntactic information, 
we define semantic information to be \emph{that syntactic
information between the system and the environment
that causally contributes to the continued existence of
the system}, i.e., to maintaining the value of the viability function. To quantify the causal contribution, we define counter-factual \textbf{intervened distributions} in which some of the syntactic information between the system and its environment is scrambled. This approach is inspired by the framework of causal interventions~\cite{pear00,ay2008information}, in which causal effects are measured by counter-factually intervening on one part of a system and then measuring the resulting changes in other parts of the system.

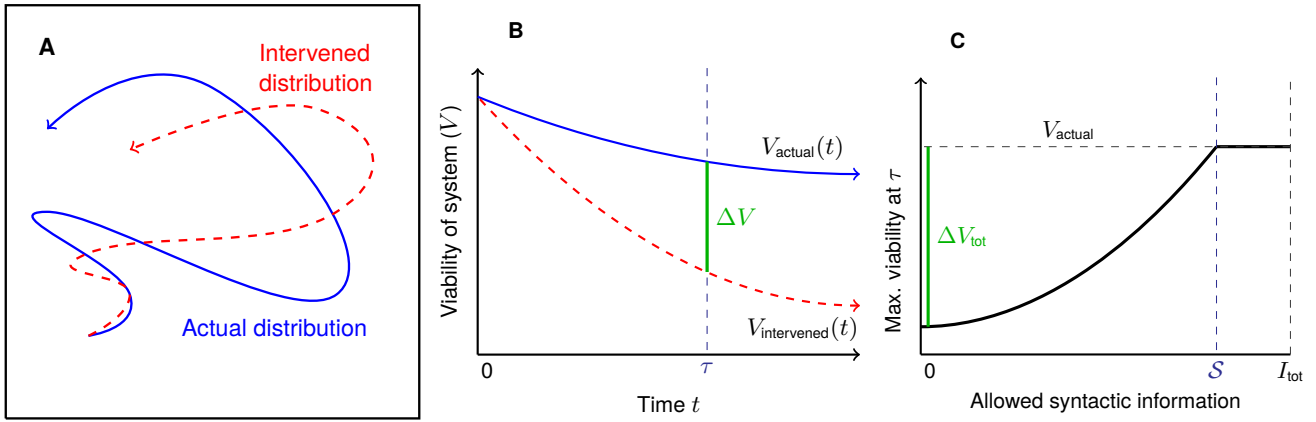
\begin{figure*}{
	\sffamily
\resizebox {0.33\textwidth} {!} {
\begin{tikzpicture}[
    scale=5,
]

\node at (0.1,0.9) {\textsf{\textbf{A}}};

\draw[thick] (0,0) -- (1,0) -- (1,1) -- (0,1) -- (0,0); 
\draw [thick,blue,->] plot [smooth, tension=1] coordinates { (0.2,0.2) (0.3, 0.3) (0.1,0.5) (0.8,0.3) (0.5,0.8) (0.1,0.7)};
\draw [thick,dashed,red,->] plot [smooth, tension=1] coordinates { (0.2,0.2) (0.3,0.3) (0.2, 0.4) (0.8,0.5) (0.75,0.75) (0.3,0.65)};
\node [blue] at (0.65,0.22) {Actual distribution};
\node [red,align=left] at (0.75,0.85) {Intervened\\distribution};
\end{tikzpicture}
}\resizebox {0.33\textwidth} {!} {
\begin{tikzpicture}[
    scale=5,
]
\node at (0.1,0.85) {\textsf{\textbf{B}}};

\def \maxy {0.75}
\def \maxs {0.9*\maxy}
\def \actualft {0.6}
\draw[thick,->] (0,0) -- (1,0); 
\draw (0.025,0) node[below] {0};
\draw[thick,->] (0,0) -- (0,\maxy); 
\node at (0.5, -0.13) {Time $t$};
\node[rotate=90] at (-0.07,0.5*\maxy) {Viability of system ($V$)};
\draw[thick,<-,blue,name path=viability1] (1.0,0.7*\maxs)  parabola (0.0,\maxs);
\node at (0.85, 0.8*\maxs) {$V_\text{actual}(t)$};
\draw[thick,<-,dashed,red, name path=viability2] (1.0,0.19*\maxs) parabola (0.0,\maxs);
\node at (0.85, 0.1*\maxs) {$V_\text{intervened}(t)$};

\draw[dashed,Blue, name path=ft] (\actualft,\maxy) -- (\actualft,0) node[below] {$\ft$};

\path [name intersections={of=viability1 and ft,by=i1}]; 
\path [name intersections={of=viability2 and ft,by=i2}]; 
 
\draw[very thick,black!30!green] (i2) -- (i1) node[midway,right,align=left] {$\Delta V$};

\end{tikzpicture}
}\resizebox {0.33\textwidth} {!} {
\begin{tikzpicture}[
    scale=5,
]
\node at (0.1,0.85) {\textsf{\textbf{C}}};

\def \maxy {0.75}
\def \maxs {0.75*\maxy}
\def \mins {0.1*\maxy}
\def \flatstart {0.8}
\draw[thick] (0,0) -- (1,0) node[below] {$I_\text{tot}$}; 
\draw (0.025,0) node[below] {0};
\draw[thick,->] (0,0) -- (0,\maxy) ; 
\node at (0.5, -0.13) {Allowed syntactic information};
\node[rotate=90] at (-0.07,0.5*\maxs) {Max. viability at $\ft$};
\draw[very thick] (0,\mins) parabola (0.8,\maxs) ;
\draw[very thick] (\flatstart,\maxs) parabola (1.0,\maxs) ;
\draw[dashed,Blue] (\flatstart,\maxy) -- (0.8,0) node[below] {$\mathcal{S}$};
\draw[dashed,darkgray] (1,\maxs) -- (0,\maxs);
\node[above] at (0.5*\flatstart, \maxs) {$V_\text{actual}$}; 

\draw[dashed,darkgray] (1,0) -- (1,\maxy);

\draw[very thick, black!30!green] (0.02,\mins) -- (0.02,\maxs) node[right,midway] {$\Delta V_\text{tot}$} ;

\end{tikzpicture}
}
}

\caption{\label{fig:fig1-schematic}Schematic illustration of our approach
to semantic information. (A) The trajectory of the actual distribution
(within the space of distribution over joint system-environment states) is in
blue. The trajectory of the intervened distribution, where some syntactic
information between the system and environment is scrambled, is in
dashed red. (B) The viability function computed for both the actual
and intervened trajectories. $\Delta V$ indicates the viability difference between actual and intervened
trajectories, at some time $\ft$. (C) Different ways of scrambling
the syntactic information lead to different values of remaining syntactic
information and different viability values. The maximum
achievable viability at time $\ft$ at each level of remaining syntactic
information specifies the information/viability curve. The viability
value of information, $\protect\Vtot$, is the total viability cost
of scrambling all syntactic information. The amount of semantic information,
$\mathcal{S}$, is the minimum level of syntactic information at which no viability is lost.
$I_{\text{tot}}$ is the total amount of syntactic information between system and environment.}
\end{figure*}

The trajectories of the actual
and intervened distributions are schematically illustrated in \cref{fig:fig1-schematic}A.
We define the \textbf{(viability) value of information} as the
difference between the system's viability after time $\ft$ under
the actual distribution, versus the system's viability after time
$\ft$ under the intervened distribution (\cref{fig:fig1-schematic}B).
A positive difference means that at least some of the syntactic information
between the system and environment plays a causal role in maintaining the
system's existence. %
The difference can also be \textit{negative}, which means that the syntactic
information decreases the system's ability to exist. This occurs if the system behaves ``pathologically'', i.e., it takes the wrong actions given available information (e.g., consider
a mutant ``anti-chemotactic'' bacteria, which senses the direction of food and then swims {away} from it).

To make things more concrete, we illustrate our approach using a few examples:
\begin{enumerate}
\item Consider a distribution over rocks (the system) and fields (the environment) over a timescale of $\ft=1$ year.  Rocks tend to stay in a low entropy state for long periods of time due to their very slow dynamics.  If we ``scramble the information'' between rocks and their environments by swapping rocks between different fields, this will not significantly change the propensity of rocks to disintegrate into (high entropy) dust after 1 year. Since the viability does not change significantly due to the intervention, the viability value of information is very low for a rock.
\item Consider a distribution over hurricanes (the system) and the summertime Caribbean ocean and atmosphere (the environment), over a timescale of $\ft = 1$ hour. Unlike a rock, a hurricane is a genuinely nonequilibrium system which is driven by free energy fluxing from the warm ocean to the cold atmosphere.  Nonetheless, if we ``scramble the information'' by placing hurricanes in new surroundings that still correspond to warm oceans and cool atmospheres, after 1 hour the intervened hurricanes' viability will be similar to that of the non-intervened hurricanes. 
Thus, like rocks, hurricanes have a low viability value of information.  
\item Consider a distribution over food-caching birds (the system) in the forest (the environment), over a timescale of $\ft = 1$ year.  Assume that at $t=0$ the birds have cached their food and stored the location of the caches in some type of neural memory. If we ``scramble the information'' by placing birds in random environments, they will not be able to locate their food and be more likely to die, thus decreasing their viability. Thus, a food-caching bird exhibits a high value of information.
\end{enumerate}

So far, we have spoken of interventions in a rather informal manner.   In order to make things rigorous, we require a formal definition of how to transform an actual distribution into an intervened distribution.
While we do not claim that there is a single best choice for defining interventions, we propose to use information-theoretic ``coarse-graining'' methods to scramble the channel between the system and environment~\cite{blackwell1953equivalent,lindley_measure_1956,shannon1958note,rauh2017coarse,nasser2017input,nasser2017characterization}. 
Importantly, such methods allow us to choose different coarse-grainings, which lets us vary the  syntactic information that is preserved under different interventions, and the resulting viability of the system at time $\ft$. 
By considering different interventions, we define a 
trade-off between 
between the amount of preserved syntactic information versus the resulting viability of
the system at time $\ft$.
This trade-off 
is formally represented by an \textbf{information/viability curve}
(\cref{fig:fig1-schematic}C), which is loosely analogous to the
rate-distortion curves  in information theory~\cite{cover_elements_2012}.

%

Note that some intervened distributions may achieve the same viability
as the actual distribution but have less syntactic information.
We call the \textbf{(viability-) optimal intervention} that intervened distribution
which achieves the same viability as the actual distribution while
preserving the smallest amount of syntactic information.
Using the optimal intervention, we define a number of interesting measures. First, by definition,
any further scrambling of the optimal intervention leads to a
change in viability of the system, relative to its actual (non-intervened) viability. We interpret this to 
mean that \emph{all syntactic
information in the optimal intervention is semantic information}.
Thus, we define the \textbf{amount of semantic information} possessed by the system as the amount of syntactic information preserved by the optimal intervention.
We show that the amount of semantic information is upper bounded by the amount of syntactic
information under the actual distribution, meaning that 
having non-zero syntactic information is a necessary, but not sufficient,
condition for having non-zero semantic information. Moreover, we can decompose the total amount of syntactic information into ``meaningful bits'' (the semantic information) and the ``meaningless bits'' (the rest), and  
define the \textbf{semantic efficiency} of the system as the ratio of the 
semantic information to the syntactic information. Semantic efficiency falls between 0 and 1, and
quantifies how much the system is ``tuned'' to only possess syntactic
information which is relevant for maintaining its existence (see also \cite{still_thermodynamic_2017}).

Because all syntactic information in the optimal intervention is semantic
information, we use the optimal intervention to define the ``content'' of the semantic information. The
\textbf{semantic content} of a particular system state \textbf{$x$} is defined as the conditional distribution (under the optimal intervention)
of the environment's states, given that the system is in state $x$. 
The semantic
content of $x$ reflects the correlations which are relevant
to maintaining the existence of the system, once all other ``meaningless''
correlations are scrambled away.
To use a previous example, the semantic content for a food-caching bird would include the conditional probabilities of different food-caching locations in the forest, given bird neural states.
By applying appropriate ``pointwise''
measures of syntactic information to the optimal intervention, we also derive measures of \textbf{pointwise semantic information} in particular system states (see \cref{sec:meaningfulinfo} for details).

As mentioned, our framework is not tied to one particular measure of syntactic information, but rather can be used to derive different kinds of semantic information from different measures of syntactic information. 
In \cref{subsec:innate-intervention}, we consider semantic information derived from the mutual information between the system and environment in the initial distribution $p(x_0,y_0)$, which defines what we call \textbf{stored
semantic information}.
Note that stored semantic information 
does not measure 
semantic information which is acquired by ongoing dynamic interactions between system and environment, which is the primary kind of semantic information discussed in the literature on autonomous agents~\cite{bickhard_autonomy_2000}.
In \cref{subsec:Observed-meaning}, 
we derive this kind of dynamically-acquired semantic information,
 which we call \textbf{observed semantic information},
from a syntactic information measure called \emph{transfer entropy}~\cite{schreiber:2000uq}.
Observed semantic information provides one quantitative definition of \textbf{observation}, as dynamically-acquired information that is used by a system to maintain its own existence, and allows us 
to distinguish observation from the mere build up of syntactic information between physical systems (as generally happens  whenever physical systems come into contact).
In \cref{subsec:Other-kinds-of}, we briefly discuss other possible choices of syntactic information measures,
which lead to other measures of semantic information.

Given recent work on the statistical physics of information processing, 
several of our measures --- including value of information and semantic efficiency --- can be given thermodynamic interpretations.  We review these connections between semantic information and statistical physics in \cref{sec:Nonequilibrium-statistical-physi}, as well as in more depth in \cref{sec:meaningfulinfo} when defining stored and observed semantic information.



To summarize, we propose a formal definition of semantic information that is applicable to any physical system. Our definition depends on the specification of a viability function, a syntactic information measure, and a way of producing interventions.  We suggest some natural ways of defining these factors, 
though we have been careful to formulate our approach in a flexible manner, allowing them to be chosen according to the needs of the researcher.  Once these factors are determined, our measures of semantic information are defined relative to choice of
\begin{enumerate}
\item the particular division of the physical world into ``the system'' and ``the environment'';
\item the timescale $\ft$;
\item the initial probability distribution over the system and environment.
\end{enumerate}
These choices specify the particular spatiotemporal scale and state-space regions that interest the researcher, and should generally be chosen in a way to be relevant to the dynamics of the system under study. For instance, if studying semantic information in human beings, one should choose timescales over which information has some effect on the probability of survival (somewhere between $\approx 100$ ms, corresponding to the fastest reaction times, to $\approx 100$ years). 
In \cref{subsec:Maximization}, we discuss how the system/environment decomposition, timescale, and initial distribution might be chosen ``objectively'', in particular so as to maximize measures of semantic information. We also discuss how this might be used to automatically identify the presence of agents in physical systems, and more generally the 
implications of our framework for an intrinsic definition of \textbf{autonomous agency} in physical systems.

The rest of the paper is laid out as follow. The next section provides a review of some relevant
aspects of nonequilibrium statistical physics. In \cref{sec:preliminaries}, we provide preliminaries concerning our notation and physical assumptions, while \cref{subsec:The-viability-function} provides a discussion of the
viability function. In \cref{sec:meaningfulinfo}, we state our formal definitions of semantic information and related concepts.
\cref{subsec:Maximization} discusses ways of automatically selecting systems, timescales, and initial distributions so as to maximize semantic information, and implications for a definition of agency.  We conclude in \cref{sec:discussion}.



%
%
%
%
%

\section{Nonequilibrium statistical physics\label{sec:Nonequilibrium-statistical-physi}}

The connection between the maintenance of low entropy and autonomous agents was first noted when considering the thermodynamics of living systems.
In particular, the fact that organisms must maintain themselves in a
low entropy state was famously proposed, in an informal manner, by Schrödinger~\cite{schr44}, as well as Brillouin~\cite{brillouin_life_1949} and others~\cite{bauer1920definition,elek_living_2012}.  This had led to an important
line of work on quantifying the entropy of various kinds of living matter
\cite{morowitz_order-disorder_1955,bonchev_symmetry_1976,davies_self-organization_2013,kempes_thermodynamic_2017_rsta}.
However, this research did not consider the role of organism-environment information
exchanges in maintaining the organism's low entropy state.

 Others have observed that organisms not only maintain a low entropy state, but constantly acquire and use information about their environment to do so~\cite{corning_thermodynamics_1998,corning_thermodynamics_1998-1,corning_control_2001,collier_complexly_1999,ben_jacob_seeking_2006,polani_information:_2009,deacon_8_2010}. Moreover, it has been suggested that natural selection can drive improvements in the mechanisms that gather and store information about the environment~\cite{krak11}. However, these proposals did not specify how to formally quantify the amount
and content of  information which contributes to the self-maintenance of any given organism.%

Recently, there has been dramatic progress in our
understanding of the physics of nonequilibrium processes which acquire, transform, and use information, as part of the development of so-called ``thermodynamics of information''~\cite{parrondo2015thermodynamics}.  It is now well understood that,
as a consequence of the Second Law of Thermodynamics, any process
that reduces the entropy of a system must incur some thermodynamic
costs.  In particular, the so-called \emph{generalized Landauer's principle}~\cite{maroney2009generalizing,turgut_relations_2009,sagawa2014thermodynamic} states that, given a system coupled to a heat bath at temperature $T$, any process that reduces the entropy of the system by $n$ bits must release at least
$n\cdot\ktlntwo$ of energy as heat (alternatively,
at most $n\cdot\ktlntwo$ of heat can be absorbed
by any process that increases entropy by $n$ bits). It has also been shown that in certain scenarios, 
heat must be generated in order to acquire
syntactic information, whether mutual information \cite{sagawa2008second,sagawa2009minimal,sagawa_nonequilibrium_2012,parrondo2015thermodynamics},
transfer entropy~\cite{prokopenko2013thermodynamic,horowitz_second-law-like_2014,cafaro_thermodynamic_2016,spinney_transfer_2016,spinney_transfer_2017},
or other measures~\cite{ito2013information,horowitz2014thermodynamics,horowitz_multipartite_2015,hartich_sensory_2016,allahverdyan_thermodynamic_2009}.

Due to these developments, nonequilibrium statistical physics now has a fully rigorous understanding of
``information-powered nonequilibrium states''~\cite{sagawa2008second,sagawa2009minimal,sagawa_generalized_2010,sagawa_nonequilibrium_2012,cao_thermodynamics_2009,esposito_stochastic_2012,horowitz2011thermodynamic,horowitz_nonequilibrium_2010,horowitz_second-law-like_2014,munakata2012entropy,munakata2013feedback,ponmurugan_generalized_2010,kim2007fluctuation,mandal2013maxwell,barato2013autonomous,koski2014experimental}, i.e., systems in 
which non-equilibrium is maintained by the ongoing exchange of information
between subsystems. The prototypical case of such situations are ``feedback-control''
processes, in which one subsystem acquires information about
another subsystem, and then uses this information to apply appropriate control
protocols so as to keep itself or the other system out of equilibrium (e.g., Maxwell's
demon~\cite{mandal2012work,barato2013autonomous,koski2014experimental}, feedback cooling~\cite{mandal2013maxwell}, etc.).
Information-powered nonequilibrium states differ from the kinds
of nonequilibrium systems traditionally considered in statistical physics,
which are driven by work reservoirs with (feedback-less) control protocols,
or by coupling to multiple thermodynamic reservoirs (e.g., Bénard
cells).

Recall that we define our viability functions as the negative entropy of the system.  As stated, 
results from 
nonequilibrium statistical physics show that both decreasing entropy (i.e., increasing viability)
and acquiring syntactic information carries thermodynamic costs, and these costs
can be related to each other. In particular, the syntactic information
that a system has about its environment will often require some
work to acquire.
However, the same information may carry an arbitrarily large benefit~\cite{seth_bit_worth}, 
for instance by indicating the location of a large source of free energy, or a danger to avoid. 
To compare the benefit and the cost of the syntactic information to the system,
below we define the \textbf{thermodynamic multiplier} as the ratio between the
viability value of the information and the amount of syntactic information. %
Having
a large thermodynamic multiplier indicates that the information that
the system has about the environment leads to a large ``bang-per-bit''
in terms of viability. As we will see, the thermodynamic multiplier is related to the semantic efficiency of a system: systems with positive value of information and high semantic efficiency tend to have larger thermodynamic multipliers.
%
%
%
%
%
%
%
%
%
%

%
%
%
%
%
%
%

\section{Preliminaries and physical setup}

\label{sec:preliminaries}

We indicate random variables by capital letters, such as $X$, and particular outcomes of random variables by corresponding lowercase letters, such as $x$.  Lower case letters $p, q, \dots$ are also  used to refer to probability distributions.  Where not clear from context, we use notation like $\p_X$ to indicate that $\p$ is a distribution of the random variable $X$.  We  also use notation like $\p_{X,Y}$ for the joint distribution of $X$ and $Y$, and $\p_{X\vert Y}$  for the conditional distribution of $X$ given $Y$. We use notation like $p_X p_Y$ to indicate product distributions, i.e., $[p_X p_Y](x,y)=p_X(x)p_Y(y)$ for all $x,y$.

We assume that the reader is familiar with the basics of information
theory~\cite{cover_elements_2012}. We  write $S(\p_X)$ for
the Shannon entropy of distribution $\p_X$, %
$I_{\p}(X;Y)$ for the mutual information between random variables
$X$ and $Y$ with joint distribution $\p_{X,Y}$, and $I_{\p}(X;Y\vert Z)$
for the conditional mutual information given joint distribution $\p_{X,Y,Z}$. 
We measure information in bits, except where noted.

In addition to the standard measures from information theory, we also
utilize a measure called \emph{transfer entropy}~\cite{schreiber:2000uq}.
 Given a distribution $p$ over a sequence of paired random variables $(X_{0},Y_{0}),(X_{1},Y_{1}),\dots,(X_{\ft},Y_{\ft})$ indexed by timestep $t\in \{0,..,\ft\}$,
the transfer entropy from $Y$ to $X$ at timestep $t$ is defined as
the conditional mutual information,
\begin{align}
\TE[p][t]=I_{p}(Y_{t};X_{t+1}\vert X_{t}) 
\label{eq:te-def}
\end{align}
Transfer entropy reflects how much knowledge of the state of $Y$
at timestep $t$ reduces uncertainty about the next state
of $X$ at the next timestep $t+1$, conditioned on knowing the state of
$X$ at timestep $t$. It thus reflects ``new information''
about $Y$ that is acquired by $X$ at time $t$.

In our analysis below, we assume that there are two coupled systems, called ``the system \sysX'' 
and ``the environment \sysY'', with state-spaces indicated by $X$ and $Y$ respectively. 
The system/environment $X\times Y$ may be isolated from the rest of the universe, or may be coupled
to one or more thermodynamic reservoirs and/or work reservoirs. 
For simplicity, we assume that the joint state
space $X\times Y$ 
is discrete and finite (in physics, such a discrete state space is often derived
by coarse-graining an underlying Hamiltonian system \cite{wehrl1978general,van1992stochastic}), though in principle our approach can also be extended to continuous state-spaces.
In some cases, 
$X\times Y$ may also represent a space of coarse-grained
macrostates rather than microstates (e.g., a vector of chemical concentrations at different spatial locations), usually under the assumption
that local equilibrium holds within each macrostate (see Appendix~\ref{app:Food-seeking-Agent-Model} for an example).

The joint system evolves dynamically from initial time $t=0$ to final time $t=\ft$. 
We assume that the decomposition into system/environment remains constant over this time (in future work, it may be interesting to consider time-inhomogeneous decompositions, e.g., for analyzing growing systems).
In our analysis of observed semantic information in \cref{subsec:Observed-meaning}, 
 we assume for simplicity that the coupled dynamics of \sysX and \sysY are stochastic, 
discrete-time and first-order Markovian. However, we do not 
 assume that that dynamics are time-homogeneous (meaning that, in principle, our framework allows for 
external driving by the work reservoir). Other kinds of dynamics (e.g., Hamiltonian dynamics, which are continuous-time and deterministic)
can also be considered, though care is needed when defining measures
like transfer entropy for continuous-time systems \cite{spinney_transfer_2017}. 

We use random variables $X_{t}$ and $Y_{t}$ to represent the state
of \sysX and \sysY at some particular time $t\ge0$, 
and random variables $X_{0..\ft}=\left\langle X_{0},\dots,X_{\ft}\right\rangle $
and $Y_{0..\ft}=\left\langle Y_{0},\dots,Y_{\ft}\right\rangle $ to
indicate entire trajectories of \sysX and \sysY from time $t=0$ to $t=\ft$.

\section{The viability function\label{subsec:The-viability-function}}

We quantify the ``level of existence'' of a given system at any given time
with a \textbf{viability function} $V$. Though several viability
functions can be considered, in this paper we define the viability
function as the negative of the Shannon entropy of the marginal distribution of system \sysX at time $\ft$,
\begin{equation}
V(\pft):=-S(\pft)=\sum_{x_{\ft}}\p(x_{\ft})\log \p(x_{\ft})\label{eq:Vmicro}
\end{equation}
If the state space of \sysX represents a set of coarse-grained macrostates, 
\cref{eq:Vmicro} should be amended
to include the contribution from ``internal entropies'' of
each macrostate (see Appendix~\ref{app:Food-seeking-Agent-Model} for an example).

There are several reasons for selecting negative entropy as the viability
function. 
First, as discussed in \cref{sec:Nonequilibrium-statistical-physi}, results in nonequilibrium statistical physics relate changes
of the Shannon entropy of a physical system to thermodynamic quantities like heat
and work~\cite{parrondo2015thermodynamics,esposito2011second,seifert2012stochastic,maroney2009generalizing,wolpert_landauer_2016a,horowitz_minimum_2017}.
These relations allow us to analyze our measures in terms of thermodynamic
costs.

The second reason we define viability as negative entropy is that entropy provides an upper bound on the amount of probability
that can be concentrated in any small subset of the
state space $X$ (for this reason, entropy has been used as a measure
of the performance of a controller~\cite{touchette_information-theoretic_2000,touchette2004information,cao_thermodynamics_2009}). For us, this is relevant because there is often a naturally-defined ``viability set''~\cite{ashby_design_1960,beer_dynamical_1996,di_paolo_autopoiesis_2005,egbert2011quantifying,agmon_exploring_2016,agmon2016structure},
which is the set of states in which the system \sysX can continue to perform self-maintenance functions.
Typically, the viability set will be a very small subset of the overall state space $X$. 
For instance, the total number of ways in which the atoms in an \emph{E. Coli} bacterium can be arranged, relative to the number
of ways they can be arranged to constitute a living \emph{E. Coli}, has
been estimated to be on the order of $2^{46,000,000}$ \cite{morowitz_order-disorder_1955}.
If the
entropy of system \sysX is large and the viability set is small, then the probability that the system state is within the viability set must be small, 
no matter where that viability
set is in $X$. Thus, maintaining low entropy is a necessary conditions for remaining within the viability set. (Appendix~\ref{app:entbound} elaborates
these points, deriving a bound between Shannon entropy and the probability of the system being within any small subset of its state space.)

At the same time,  negative entropy 
 may have some disadvantages as a viability function. 
Most obviously, a distribution can have low entropy but
still assign a low probability to being in a particular viability
set. In addition, a system that maintains low entropy over time does not necessarily
``maintain its identity'' (e.g.,
both a rhinoceros and a human have low entropy). Whether this is an advantage or a drawback of the measure depends partly on how the notion of ``self-maintenance'' is conceptualized.

There are other ways to define the viability function, some of which address these potential disadvantages of using negative entropy.
Given a particular viability set $\Aset \subseteq X$, a natural definition of the viability function is the probability that the system's state is in the viability set, $\p(X_\ft \in\Aset)$. However,
this definition requires the viability set to be specified, and in many
scenarios we might know that there is a viability set but not be able to 
specify it precisely. To use a previous example, identifying the viability set of an \emph{E. Coli} is 
an incredibly challenging problem~\cite{morowitz_order-disorder_1955}.

Alternatively, it is often stated 
that self-maintaining systems must remain out of
thermodynamic equilibrium~\cite{collier_complexly_1999,bickhard_autonomy_2000,ruiz-mirazo_basic_2004}.
This suggests defining the viability function in a way that captures the ``distance
from equilibrium'' of system \sysX. One such measure is the Kullback-Leibler divergence (in nats) between the actual distribution over $X_\ft$ and the equilibrium distribution of \sysX at time $\ft$, indicated here by $\pi_{X_\ft}$,
\begin{equation}
\KL(\pft \Vert \pi_{X_\ft})\label{eq:kl00}
\end{equation}
This viability function, which is sometimes called ``exergy''
or ``availability'' in the literature \cite{schlogl1985thermodynamic,schneider_life_1994},   
has a natural physical interpretation~\cite{esposito2011second}:
if the system were separated from environment \sysY and coupled to a single heat bath at temperature $T$, then up to
$k_{B}T\cdot\KL(\pft \Vert \pi_{X_\ft})$ work could be extracted by bringing the system from $\pft$ to
$\pi_{X_\ft}$.

Unfortunately, there are difficulties in using \cref{eq:kl00} as the
viability function in the general case. In statistical physics, the
equilibrium distribution is defined as a stationary distribution in which
all probability fluxes vanish. Since the system
\sysX is open (it is coupled to the environment \sysY, and possibly multiple 
thermodynamic reservoirs), such an equilibrium distribution
will not exist in the general case, and \cref{eq:kl00} may be undefined. For
instance, a Bénard cell, a well-known nonequilibrium system which is
coupled to both hot and cold thermal reservoirs~\cite{chandrasekhar1961hydrodynamic},
will evolve to a \emph{non-equilibrium} stationary distribution, in which
probability fluxes do not vanish. While it is certainly true that a 
Bénard cell is out of thermodynamic equilibrium, one cannot quantify
``how far'' from equilibrium it is by using \cref{eq:kl00}.


In principle, it is possible to quantify the ``amount of non-equilibrium'' without making reference to an equilibrium distribution, in particular by
measuring the amount of probability flux in a system
(e.g., instantaneous entropy production
\cite{esposito2010three,van2010three} or the norm of the probability
fluxes \cite{zia2007probability,platini2011measure}). However, 
there is not necessarily a clear relationship between the amount of probability flux and the capacity of a system to carry out
self-maintenance functions~\cite{baiesi2018life}.
We leave exploration of these alternative  viability functions for future work.

It is important to re-emphasize that, in our framework, the viability function is exogenously determined by the scientist analyzing the system, rather than being a purely endogenous characteristic of the system. At first glance, our approach may appear to suffer some of the same problems as do approaches that define semantic information in terms of an exogenously-specified utility function (see the discussion in the Introduction).  However, there are important differences between a utility function and a viability function. First, we require that a viability function is well-defined for \textit{any} physical system, whether a rock, a human, a city, a galaxy; utility functions, on the other hand, are generally scenario-specific and far from universal. Furthermore, given an agent with an exogenously-defined utility function operating in a time-extended scenario, maintaining existence is almost always a necessary (though usually implicit) condition for high utility. A reasonably-chosen viability function should capture this minimal, universal component of nearly all utility functions. Finally, unlike utility functions, in principle it may be possible to derive the viability function in some objective way (e.g., in terms of the attractor landscape of the coupled system-environment dynamics~\cite{ashby_design_1960,agmon2016structure}). 

\section{Semantic information via interventions}

\label{sec:meaningfulinfo}

As described above, we quantify semantic information in terms of the amount of syntactic
information which contributes to the ability of the system to
continue existing. 

We use the term \textbf{actual distribution} 
to refer to the original, unintervened distribution of trajectories
of the joint system-environment over time $t=0$ to $t=\ft$, which will usually be indicated with the symbol  $p$. 
Our goal is to quantify
how much semantic information the system has about the environment under
the actual distribution. To do this, we define a set of counter-factual \textbf{intervened
distributions} over trajectories, which are similar to the actual distribution except that some of syntactic information between system and environment is scrambled, and which 
will usually be indicated with some variant of the symbol $\hat{p}$. 
%
%
%
We define measures of semantic information by analyzing how the viability of the system at time $\ft$ 
changes between the actual and the intervened distributions.
%
%
%
%
%
%
%


Information theory provides many different measures of syntactic information
between the system and environment, each of which requires a special type of intervention, and each of which gives rise to a particular set of semantic information measures. 
In this paper, 
we focus on two types of syntactic information. In \cref{subsec:innate-intervention}, we consider 
\textbf{stored
semantic information}, which is defined by scrambling the mutual information between system and environment in the actual initial distribution $p_{X_0,Y_0}$, while leaving the dynamics  unchanged.
In \cref{subsec:Observed-meaning}, we instead consider \textbf{observed semantic information}, which is defined via a ``dynamic'' intervention in which we keep the initial distribution the same but change the
dynamics so as to scramble the transfer entropy from the environment to
the system. Observed semantic information identifies semantic information that is acquired by dynamic interactions between the system and environment, rather than present in the initial mutual information.
An example of observed semantic information is exhibited by a chemotactic bacterium, which
makes ongoing measurements of the direction of food in its environment, and then uses this information to move toward food. 
In \cref{subsec:Other-kinds-of}, we briefly discuss other possible measures of semantic information.


\subsection{Stored semantic information\label{subsec:innate-intervention}}
\subsubsection{Overview}

\textbf{Stored semantic information} is derived from the mutual information between system and environment at time $t=0$.  This mutual information can be written as
\begin{align}
\MI[p][0]= \sum_{x_0, y_0} p(x_0, y_0) \log \frac{p(x_0, y_0)}{p(x_0) p(y_0)} \,.
\label{eq:miformula}
\end{align}
Mutual information achieves its minimum value of  0 if and only if $X_0$ and $Y_0$ are statistically-independent under $p$, i.e., when $p_{X_{0},Y_{0}} = p_{X_{0}}p_{Y_{0}}$.
%
Thus, we first consider an intervention that destroys all mutual information by transforming
the actual initial distribution $p_{X_{0},Y_{0}}$ to the product initial distribution,
\begin{align}
\label{eq:scramble}
p_{X_0,Y_0} \mapsto \pfull_{X_{0},Y_{0}}:=p_{X_{0}}p_{Y_{0}} \,. 
\end{align}
(We use the superscript $\text{``full''}$ to indicate that this is a ``full scrambling'' of the mutual information.)%

To compute the \textbf{viability value} of stored semantic information at
$t=0$, we run the coupled system-environment dynamics starting from both the actual
initial distribution $p_{X_{0},Y_{0}}$ and the intervened initial
distribution $\pfull_{X_{0},Y_{0}}$, and then measure
the difference in the viability of the system at time $\ft$,
\begin{align}
\VtotS & :=V(\pft)-V(\pfft)\label{eq:value}
\end{align}
For
the particular viability function we are considering (negative entropy), 
the viability value is
\begin{equation}
\VtotS=S(\pfft)-S(\pft)\label{eq:vtotent}
\end{equation}

\cref{eq:value} measures the difference of  viability under the ``full
scrambling'', but does not specify which part of the mutual
information actually causes this difference.  To illustrate this issue, 
consider 
a system in an environment
where food can be in one of two locations with 50\% probability
each, and the system starts at $t=0$ with perfect information about the food location. Imagine that system's viability depends upon it finding and eating the food.
Now suppose that the system also has 1000 bits of mutual
information about the state of the environment which does not contribute
in any way to the system's viability. 
In this case, the initial mutual information will be 1001 bits, though only 1 bit (the location of the
food) is ``meaningful'' to the system, in that it affects the system's ability
to maintain high viability.

In order to find that part of the mutual information which is meaningful, we define an entire set of ``partial'' interventions (rather than just considering considering the single ``full'' intervention mentioned above). We then find the 
partial intervention which destroys the most syntactic information while leaving
the viability unchanged, which we call the \textbf{(viability-) optimal intervention}. The optimal intervention specifies which part of the mutual information is meaningless, in that it can be scrambled  without affecting viability, and which part is meaningful, in the sense that it must be preserved in order to achieve the actual viability value.
For the example mentioned in the previous paragraph, the viability-optimal intervention would preserve the 1 bit of information concerning the location of the food, while scrambling away the remaining 1000 bits.


Each partial interventions in the set of possible partial interventions is induced by a particular ``coarse-graining function''.  First, consider the actual conditional probability of system given environment at $t=0$, $p_{X_0\vert Y_0}$, as a communication channel over which the system acquires information from its environment. 
To define each partial intervention, we   
 coarse-grain this communication channel $p_{X_0\vert Y_0}$ using a 
coarse-graining function $\cg(y)$, which specifies which distinctions the system can make about the environment.  Formally, the intervened channel from $Y_0$ to $X_0$ induced by $\cg$, indicated as $\pcg_{X_0\vert Y_0}$, is taken to be the  actual conditional probability of system states $X_0$ given coarse-grained environments $\cg(Y_0)$,
\begin{align}
\label{eq:condchanneldef}
\pcg(x_0\vert y_0) \!:=\! 
 p(x_0\vert  \cg(y_0)) \!=\! \frac{\sum_{y_0':\cg(y_0')=\cg(y_0)} p(x_0,y_0')}{\sum_{y_0':\cg(y_0')=\cg(y_0)} p(y_0')} \,.
\end{align}
We then define the intervened joint distribution at $t=0$ as $\pcg_{X_0,Y_0} := \pcg_{X_0\vert Y_0} p_{Y_0}$.
Under the intervened distribution $\pcg_{X_0,Y_0}$, $X_0$ is conditionally independent of $Y_0$ given $\cg(Y_0)$, and any two states of the environment $y_0$ and $y_0'$ which have $\cg(y_0)=\cg(y_0')$ will be indistinguishable from the point of view of the system.   Said differently,  $X_0$ will only have information about $\cg(Y_0)$, not $Y_0$ itself, and it can be verified that $I_{\pcg}(X_0;Y_0) = I_p(X_0;\cg(Y_0))$.  
In the information-theory literature, the coarse-grained channel $\pcg_{X_0\vert Y_0}$ is sometimes called a ``Markov approximation'' of the actual channel $p_{X_0\vert Y_0}$~\cite{rauh2017coarse},
which is itself a special case of so-called ``channel pre-garbling'' or ``channel input-degradation''~\cite{rauh2017coarse,nasser2017input,nasser2017characterization}. Pre-garbling is a principled way to destroy part of the information flowing across a channel, and has important operationalizations in terms of coding and game theory~\cite{nasser2017input}.

So far we have left unspecified how the coarse-graining function $\cg$ is chosen.  In fact, one can choose  different $\cg$, in this way inducing different partial interventions. 
The ``most conservative'' intervention corresponds to any $\cg$ which is a one-to-one function of $Y$,
such as the identity map $\cg(y)=y$. In this case, one can use \cref{eq:condchanneldef} to verify that the intervened
channel from $Y_0$ to $X_0$ will be same as the actual channel, and the intervention will have no effect. 
The ``least conservative''
intervention occurs when $\cg$ is a constant function, such as $\cg(y)=0$. In this case, the intervened distribution 
will be the ``full scrambling'' of
\cref{eq:scramble}, for which $I_{\pcg}(X_{0};Y_{0})=0$. 
We use $\Cg$ to indicate the set of all possible coarse-graining functions 
 (without loss of generality, we can assume that each element of this set is $\cg : Y \rightarrow Y$).

%
%

We are now ready to define our remaining measures of stored semantic information.
 We first
 define the \textbf{information/viability curve} as the maximal
achievable viability at time $\ft$ under any possible intervention,
\begin{align*}
\DS(R):=\max_{\cg \in \Cg}V(\pcgft)\quad\text{s.t.}\quad\MI[\pcg][0]=R \,,
\end{align*}
where  $R$ indicates the amount of mutual information that is preserved.
(Note that $\DS(R)$ is undefined for values of $R$ when there is no function $\cg$ such that $\MI[\pcg][0]=R$.)
$\DS(R)$ is
the curve schematically diagrammed in \cref{fig:fig1-schematic}C.

We define the \textbf{(viability-) optimal intervention} $\pcs_{X_0,Y_0}$ as the intervention  that achieves the same viability value as the
actual distribution while having the smallest amount of syntactic information, %
\begin{align}
\pcs_{X_{0},Y_{0}} \in\argmin_{\pcg : \cg \in \Cg }\MI[\pcg][0] \quad \text{s.t.} \quad V(\pcgft)=V(\pft) \,.
\label{eq:PCS}
\end{align}
By definition, any further scrambling of $\pcs_{X_0,Y_0}$ would change system viability, meaning
that in $\pcs_{X_0,Y_0}$ all 
remaining 
mutual information is meaningful. Therefore, we define the
\textbf{amount of stored semantic information} as the mutual information
in the optimal intervention,
\begin{equation}
\SS:=\MI[\pcs][0]\label{eq:storedmeaninfo}
\end{equation}
While the value of information $\VtotS$ can be positive
or negative, the amount of stored semantic information is always non-negative.
Moreover, stored semantic information reflects the number of
bits that play a causal role in determining the viability of the system
at time $\ft$, regardless in whether they cause it to change positively
or negatively.%

%
%
%
%
%

Since the actual distribution $p_{X_0,Y_0}$ is part of the domain of the minimization in \cref{eq:PCS} (it corresponds to any $\cg$ which is one-to-one), 
%
the amount of stored semantic information $\MI[\pcs][0]$ must be less
than the actual mutual information $\MI[p][0]$. We
define the \textbf{semantic efficiency} as the ratio of the stored semantic
information to the overall syntactic information,
\begin{align}
\label{eq:semeff}
\pS:=\frac{\SS}{\MI[][0]}\in[0,1]
\end{align}
Semantic efficiency measures what portion of the initial mutual information
between the system and environment causally contributes to the
viability of the system at time $\ft$. 

%
%
%

%
%
%
%
%

\subsubsection{Pointwise Measures}

As mentioned, the optimal intervention  only contains semantic information, i.e., only information which affects the viability of the system at time $\ft$. We use this to define the \textbf{pointwise semantic information} of individual states of the system and environment in terms of ``pointwise'' measures of 
mutual information~\cite{bouma2009normalized} under $\pcs$,
\begin{equation}
\sS(x_0;y_0) := \log\frac{\pcs(x_{0},y_{0})}{\pcs(x_{0})\pcs(y_{0})}\,.\label{eq:pointwisemeaninfo}
\end{equation}
We similarly define the \textbf{specific semantic information} in system state $x_{0}$ as
the ``specific information''\cite{deweese1999measure} about $Y$ given $x_{0}$,
\begin{equation}
\sS(x_0;Y_0) = \sum_{y_{0}}\pcs(y_{0}\vert x_{0})\log\frac{\pcs(y_{0}\vert x_{0})}{\pcs(y_{0})}\,.\label{eq:specificMI}
\end{equation}
{These measures quantify the extent to which a system state $x_0$, and a system-environment state $x_0,y_0$, carry correlations which causally affect the system's viability at $t=\ft$.}
Note that the specific semantic information, \cref{eq:specificMI}, and
overall stored semantic information, \cref{eq:storedmeaninfo}, are
expectations of the pointwise semantic information, \cref{eq:pointwisemeaninfo}.

Finally, we define the \textbf{semantic content} of
system state $x_{0}$ as the conditional distribution $\pcs(y_{0}\vert x_{0})$ over all $y_0\in Y$. 
The semantic content of $x_0$ reflects the precise set of correlations between $x_{0}$ and the 
environment at $t=0$ that causally affect the system's viability at time $\ft$.

It is important to note that the optimal intervention may not be unique, i.e., there might be multiple minimizers of \cref{eq:PCS}.  In case there are multiple optimal interventions, each optimal intervention will have its own measures of semantic content, and its own measures of pointwise and specific semantic information.
The non-uniqueness of the optimal intervention, if it occurs, indicates that the system possesses multiple \emph{redundant} sources of semantic information,  any one of which is sufficient to achieve the actual viability value at time $\ft$. A prototypical example is when the system has information about multiple sources of food which all provide the same viability benefit, and where the system can access at most one food source during $t\in[0,\ft]$.

\subsubsection{Thermodynamics}
\label{subsec:thermo}


In this section, we use ideas from statistical physics to define the \textbf{thermodynamic multiplier} of stored semantic information. This measure compares the physical costs to the benefits of system-environment mutual information. 


We begin with a simple illustrative example. Imagine a system coupled to a heat bath at temperature $T$, as well as an environment which contains a source of $10^{6}$ J of free energy (e.g., a hamburger) in one of two locations
(A or B), with 50\% probability each. Assume that the system only has time to move to only one of these locations during the interval $t\in[0,\ft]$.
We now consider two scenarios.  In the first, the system  initially has 1 bit of information about the location of the hamburger, which will 
generally cost at least $k_B T \ln 2$ of work to acquire.
The system can use this information to
move to the hamburger's location and then extract $10^{6}$ J of free energy.
In the second scenario, the system never acquires the 1 bit of information about the hamburger location, and instead starts from the ``fully scrambled'' distribution $\pfull_{X_0,Y_0} = p_{X_0} p_{Y_0}$ (\cref{eq:scramble}). By not acquiring the 1 bit of information, the system can save  $k_B T \ln 2$ of work, which could  
 be  used at time $\ft$ to decrease its entropy (i.e., increase its viability) by 1 bit.  However, because the system has no information about the hamburger location, it only finds the hamburger 50\% of the time, thereby missing out on $0.5 \times 10^{6}$ J of free energy on average. This amount of lost free energy could have been used to decrease the system's entropy by $\frac{0.5 \times 10^{6}}{k_B T \ln2}$ bits at time $t=\ft$. At typical temperatures, $\frac{0.5 \times 10^{6}}{k_B T \ln2} \gg 1$, meaning that 
 the benefit of having the bit of information about the hamburger location  far outweighs the cost of acquiring that bit.

To make this argument formal, imagine a physical ``measurement'' process that transforms the fully-scrambled system-environment distribution $\pfull_{X_0,Y_0} = p_{X_0}p_{Y_0}$ to the actual joint distribution $p_{X_0,Y_0}$.
Assume that during the course of this process, the interaction energy between \sysX and \sysY is negligible and that a heat bath at temperature $T$ is available. The minimum amount of work required by any such measurement process~\cite{sagawa2009minimal,parrondo2015thermodynamics} is $k_B T \ln 2$ times the change of system-environment entropy in bits, 
$\Delta S = [S(p_{X_0}) + S(p_{Y_0})] - S(p_{X_0,Y_0}) = I_p(X_0;Y_0)$.  We take this minimum work,
\begin{align}
W_\text{min} =k_{B}T \ln 2 \cdot I_{p}(X_{0};Y_{0}) \,,
\label{eq:workcost}
\end{align}
to be the cost of acquiring the mutual information.  If
%
this work were not spent acquiring the initial mutual information, it could have been used at time $\ft$ to 
decrease  the entropy of the system, and thereby increase its viability, by $I_{p}(X_{0};Y_{0})$ (again ignoring energetic considerations).






%
%
%
%
%
%
%
%
%
%
%
%
%

The benefit of the mutual information is quantified by the viability value  $\VtotS$, which reflects the difference in entropy at time $t=\ft$ when the system is started in its actual initial distribution $p_{X_0,Y_0}$ versus the  fully-scrambled initial distribution $\pfull_{X_0,Y_0} = p_{X_0}p_{Y_0}$, as in \cref{eq:vtotent}. 

Combining, we define 
the \textbf{thermodynamic multiplier}  of stored
semantic information, $\nS$, as the benefit/cost ratio of the mutual information,\footnote{Interestingly, the thermodynamic multiplier is related to an information-theoretic measure of efficiency of closed-loop control suggested in \cite[Eq. 54]{touchette2004information}.} 
\begin{equation}
\nS=\frac{\VtotS}{I_{p}(X_{0};Y_{0})}=\frac{S(\pfft)-S(\pft)}{I_{p}(X_{0};Y_{0})} \,.
\label{eq:efficiency}
\end{equation}
The thermodynamic multiplier quantifies the ``bang-per-bit'' that the syntactic information provides to the system, and provides a way to compare the ability of different systems to use information to maintain their viability high.
$\nS> 1$ means that the benefit of the information outweighs its cost.
The thermodynamic multiplier can also be related to semantic efficiency, \cref{eq:semeff}, via 
$$\nS = \pS \frac{\VtotS}{\SS} \,.$$
If the value of information is positive, then having a low semantic efficiency $\pS$ translates into having a low thermodynamic multiplier.  
Thus, there is a connection between ``paying attention to the right information'', as measured by semantic efficiency, and being thermodynamically efficient.

It is important to emphasize that we do not claim that the system \emph{actually} spends $k_{B}T \ln 2 \cdot I_{p}(X_{0};Y_{0})$ of work to acquire the mutual information in $\p_{X_0,Y_0}$. 
The actual cost could be larger, or it could be paid by the environment \sysY rather than the system, or by an external agent that prepares the joint initial condition of \sysX and \sysY, etc. 
Instead, the above analysis provides a way to compare the thermodynamic cost of acquiring the initial mutual information to the viability benefit of that mutual information. In situations where the actual cost of measurements performed by a system can be quantified (e.g., by counting the number of used ATPs), one could define the thermodynamic multiplier in terms of this actual cost.

Finally, we also emphasize that we ignore all energetic constraints in the above operationalization of the thermodynamic multiplier, in part by assuming a negligible interaction energy between system and environment. %
We have similarly ignored all energetic consequences in our analysis of interventions, as described above. It is not clear whether this approach is always justified. For instance, imagine that the system and environment have a large interaction energy at $t=0$. 
In this case, a ``measurement process'' that performs the transformation  $p_{X_0}p_{Y_0} \mapsto p_{X_0,Y_0}$ --- or alternatively an ``intervention process'' that performs the full scrambling $p_{X_0,Y_0} \mapsto p_{X_0}p_{Y_0}$ --- may involve a very large (positive or negative) change in expected energy.
Assuming the system-environment Hamiltonian is specified, one may consider defining a thermodynamic multiplier that takes into account changes in expected energy. 
Furthermore, one may also consider defining interventions in a way that obeys energetic considerations, 
so that interventions scramble information without injecting or extracting a large amount of energy into the system and environment.
Exploring such extensions remains for future work.

\begin{figure*}
\centering
\includegraphics[width=5in]{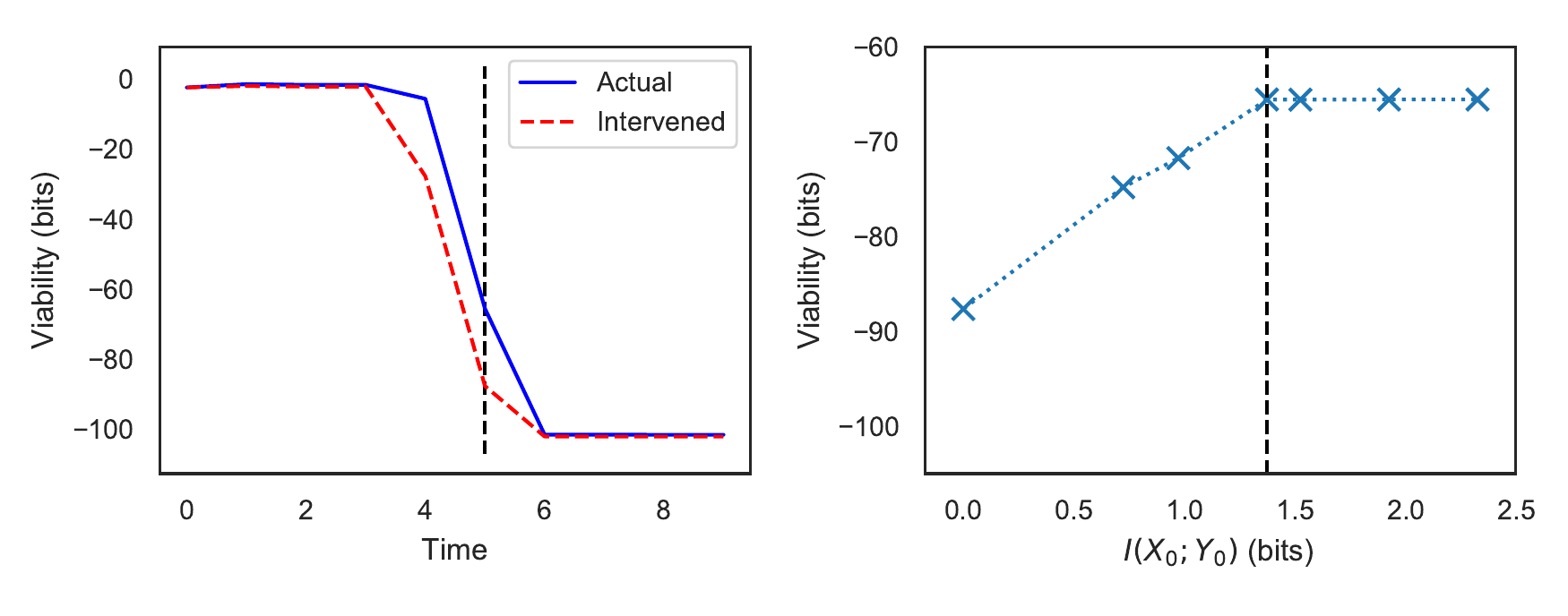}

\caption{Illustration of our approach using a simple model of a food-seeking
agent. 
On the left, we plot viability values over time under both the actual
and (fully scrambled) intervened distributions. The vertical dashed line corresponds to our timescale
of interest ($\ft=5$ timesteps). On the right, we plot the
information/viability curve for $\ft=5$ ($\times$'s are actual
points on the curve, dashed line is interpolation). 
The vertical dashed line indicates the amount of stored semantic information. See text for details.
\label{fig:stored-agent-fig}
}
\end{figure*}

\subsubsection{Example: Food-Seeking Agent}
\label{sec:stored-example}

We demonstrate our framework using a simple model of a food-seeking
agent. In this model, the environment \sysY contains 
food in one of 5 locations (initially uniformly distributed).
The agent \sysX can also be located in one of these 5 locations, and has internal information about the location of the food (i.e., its ``target'').
The agent always begins in location 3 (the middle of the world).  Under the actual initial distribution, the agent has exact information about the location of the food.
In each timestep, the agent moves towards its target and if it ever finds itself within 1 location of the food, it eats the food.  If the agent does not eat food for a certain number of timesteps, it enters a high-entropy ``death'' macrostate, which it can only exit with an extremely small probability (on the order of $\approx 10^{-34}$). 

\cref{fig:stored-agent-fig} shows the results for timescale $\ft=5$. The initial mutual information is $\log_{2}5\approx2.32$
bits, corresponding to the 5 possible locations of the food. However, 
the total amount of stored semantic information is only $\approx1.37$
bits, giving a semantic efficiency of $\pS \approx 0.6$. This occurs because
if the food is initially in locations $\{2,3,4\}$, the agent is
close enough to eat it immediately. From the point of view of the agent, differences between these three locations are ``meaningless'' and can be scrambled with no loss of viability. 
Formally, the (unique) optimal intervention $\pcs$ is induced by the following coarse-graining function,
\begin{align*}
\cg(y_0) = \begin{cases}
1 & \text{if $y_0=1$} \\
3 & \text{if $y_0 \in \{2,3,4\}$}\\
5 & \text{if $y_0=5$} \\
\end{cases}
\end{align*}
which is neither one-to-one nor a constant function (thus, it is a strictly partial intervention).
%
The value of information is $\VtotS\approx22.1$ bits, giving a thermodynamic multiplier of $\nS \approx9.5$
(the food is ``worth'' about 9.5 times more than the possible cost of acquiring information about its location).

In Appendix~\ref{app:Food-seeking-Agent-Model}, we describe this model in detail, as well as a variation in which the system moves \emph{away} from
food rather than towards it, and thus has negative value of information.  A Python implementation can be found at \url{https://github.com/artemyk/semantic_information/}.

\subsection{Observed semantic information\label{subsec:Observed-meaning}}



To identify dynamically-acquired semantic information, which we call \textbf{observed semantic information},
 we define  interventions in which we perturb 
%
the dynamic flow of syntactic information
from environment to system, without modifying the initial system-environment distribution. While there are many ways of quantifying such information flow, here we focus
on a widely-used measure called transfer entropy~\cite{schreiber:2000uq}. Transfer entropy has several attractive
features: it is directed (the transfer entropy from environment to
system is not necessarily the same as the transfer entropy from system
to environment), it captures common intuitions about information flow,
and it has undergone extensive study, including in nonequilibrium
statistical physics \cite{prokopenko2013thermodynamic,horowitz_second-law-like_2014,cafaro_thermodynamic_2016,spinney_transfer_2016,spinney_transfer_2017}.



Observed semantic information can be illustrated with the following example. Imagine a system coupled to an environment in which the food can be in one of two locations (A or B), each of which occurs with 50\% probability. At $t=0$, the system has no information about the location of the food, but the dynamics are such that it acquires and internally stores this location in transitioning from $t=0$ to $t=1$.  If we intervene and ``fully'' scramble the transfer entropy, then in transitioning from $t=0$ to $t=1$ the system would find itself ``measuring'' location A and B with 50\% probability each, independently of the actual food location. Thus, if the system used its measurements to move toward food, it would find itself finding food with only 50\% probability, and its viability would suffer. In this case, the transfer entropy from environment to system would contain observed semantic information.

Our approach is formally and conceptually similar to the one used to define stored semantic information (\cref{subsec:innate-intervention}), and we proceed in a more cursory manner.

The transfer entropy from $Y$ to $X$ over $t\in[1..\ft]$ under the actual distribution can be expressed as a sum of conditional mutual information terms (see \cref{eq:te-def}), 
\begin{align}
\sum_{t=0}^{\ft-1}\TE[p] & =\sum_{t=0}^{\ft-1} I_p(X_{t+1}; Y_t \vert X_t) \,.
\label{eq:te-condmi-main}
\end{align}
Note that the overall stochastic dynamics of the system and environment at time $t$ can be written as
$p_{X_{t+1},Y_{t+1}\vert X_t, Y_t} = p_{X_{t+1} \vert X_t, Y_t} \; p_{Y_{t+1} \vert X_t, Y_t, X_{t+1}}$, where 
$p_{X_{t+1} \vert X_t, Y_t}$ represents the response of the system to the previous state of itself and the environment, while $p_{Y_{t+1} \vert X_t, Y_t, X_{t+1}}$ represents the response of the environment to the previous state of itself and the system, as well as the current state of the system.
Observe that the conditional mutual information at time $t$ depends only on $p_{X_{t+1} \vert X_t, Y_t}$, not on $p_{Y_{t+1} \vert X_t, Y_t, X_{t+1}}$.  Thus, we define a set of partial interventions in which we partially-scramble the conditional distribution $p_{X_{t+1} \vert X_t, Y_t}$,  while keeping the conditional distribution $p_{Y_{t+1}\vert X_t, Y_t, X_{t+1}}$ undistributed.
This insures that %
our interventions  
only perturb the information flow from the environment to the system, and not vice versa.\footnote{We assume that the conditional distribution $p_{Y_{t+1}\vert X_t, Y_t, X_{t+1}}$ is fully specified. This is always the case if the conditional 
distribution $p_{X_{t+1}\vert X_{t},Y_{t}}$ is strictly positive for
all $x_{t}$, $y_{t}$, $x_{t+1}$, since then $p(y_{t+1}\vert x_{t},y_{t},x_{t+1})=\frac{p(x_{t+1},y_{t+1}\vert x_{t},y_{t})}{p(x_{t+1}\vert x_{t},y_{t})}$. If $p_{X_{t+1}\vert X_{t},Y_{t}}$ is not strictly positive, then $p_{Y_{t+1}\vert X_t, Y_t, X_{t+1}}$ has to be explicitly
provided, e.g., by specifying the joint stochastic dynamics via an appropriate Bayesian network.}

We now define our intervention procedure formally.  
As mentioned, the conditional distribution $p_{X_{t+1}\vert X_t, Y_t}$  specifies how information flows from the environment to the system at time $t$. 
Each partial intervention is defined by using a {coarse-graining function} $\cg(y)$, which is used to produce an intervened ``coarse-grained'' version of this conditional distribution at all times $t$.
The intervened conditional distribution induced by $\cg$ at time $t$, indicated as $\pcg_{X_{t+1}\vert X_t, Y_t}$, is defined to be the same as the conditional distribution of $X_{t+1}$ given $X_t$ and the coarse-grained environment $\cg(Y_t)$, 
\begin{align}
\pcg(x_{t+1}\vert x_t , y_t) & :=  \pcg(x_{t+1}\vert x_t , \cg(y_t)) \label{eq:chdef2}
\\
& = \frac{\sum_{y_t':\cg(y_t')=\cg(y_t)} p(x_{t+1}\vert x_t, y_t') \; \pcg(x_t, y_t')}{\sum_{y_t':\cg(y_t')=\cg(y_t)} \pcg(x_t, y'_t)} \,.
\end{align}
Note that this definition depends on both the actual dynamics, $p_{X_{t+1}\vert X_t,Y_t}$ and on the intervened system-environment distribution at time $t$, $\pcg_{X_t,Y_t}$.
Under the intervened distribution, $X_{t+1}$ is guaranteed to only have conditional information about $\cg(Y_t)$, not $Y_t$ itself; formally, one can verify that $I_{\pcg}(X_{t+1};Y_{t}\vert X_t) = I_p(X_{t+1};\cg(Y_t)\vert X_t)$.
These definitions are largely analogous to the ones defined for stored semantic information, and the reader should consult that section for more motivation of such coarse-graining procedures.


Under the intervened distribution, the joint system-environment dynamics at time $t$ are  computed as $\pcg_{X_{t+1}, Y_{t+1}\vert X_t, Y_t}:=\pcg_{X_{t+1}\vert X_t, Y_t} \; p_{Y_{t+1}\vert X_t, Y_t, X_{t+1}}$.
Then, the overall intervened dynamical trajectory from time $t=0$ to $t=\ft$, indicated by $\pcg_{X_{0..\ft},Y_{0..\ft}}$, is computed via the following iterative procedure:
\begin{enumerate}
	\item At $t = 0$, the intervened system-environment distribution is equal to the actual one, $\pcg_{X_0,Y_0} = p_{X_0,Y_0}$.
	\item Using $\pcg_{X_t,Y_t}$ and the above definitions, compute $\pcg_{X_{t+1}, Y_{t+1}\vert X_t, Y_t}$.
	\item Using $\pcg_{X_{t+1}, Y_{t+1}\vert X_t, Y_t}$, update 
	$\pcg_{X_{0..t},Y_{0..t}}$ to $\pcg_{X_{0..t+1},Y_{0..t+1}}$.
	\item Set $t \leftarrow t+1$ and repeat the above steps if $t < \ft$.
\end{enumerate}

We define $\Cg$ to be set of all possible coarse-graining functions.
By choosing different coarse-graining functions $\cg \in \Cg$, we can produce different partial interventions. One can verify from \cref{eq:chdef2} that the intervened distribution $\pcg_{X_{0..\ft},Y_{0..\ft}}$ will  equal to the actual
$p_{X_{0..\ft},Y_{0..\ft}}$  whenever $\cg$ is a one-to-one
function. 
When $\cg$ is a constant function, the intervened distribution will be a ``fully scrambled'' one, in which $X_{t+1}$ is conditionally independent of $Y_t$ given $X_t$ for all times $t$,
\begin{align}
\pfull_{X_{t+1}\vert X_t, Y_t} = \pcg_{X_{t+1}\vert X_t} \,.
\label{eq:obsscramble}
\end{align}
In this case, the transfer entropy at every time step will vanish.



We are now ready to define our measures of observed semantic information, which are analogous to the definition in \cref{subsec:innate-intervention}, but now defined for transfer entropy rather than initial mutual information. The \textbf{viability value
of transfer entropy} is the difference in viability at time $\ft$
between the actual distribution and the fully scrambled distribution,
\begin{align}
\VtotO & =V(\pft)-V(\pfft) \,,\label{eq:value-1}
\end{align}
where $\pfft$ is the distribution over $X$ at time $\ft$ induced by the fully scrambled intervention. 
The viability value measures the overall impact of scrambling all transfer entropy on viability.  
We define  \textbf{information/viability curve} as the
maximal achievable viability for any given level of preserved transfer entropy,
\begin{align*}
\DO(R):= \max_{\cg} \enskip & V(\pcgft)  \\
\text{s.t.} \quad & \sum_{t=0}^{\ft-1}\TE[\pcg]=R \,.
\end{align*}
The \textbf{(viability-) optimal intervention} $\pcs_{X_{0..\ft},Y_{0..\ft}}$ is defined as the intervened
distribution that achieves the same viability value as
the actual distribution while having the smallest amount of transfer entropy,
\begin{align}
\begin{aligned}
\pcs_{X_{0..\ft},Y_{0..\ft}} \in \argmin_{\pcg:\cg \in \Cg} \enskip & \sum_{t=0}^{\ft-1}\TE[\pcg] 
\\ \text{s.t.} \quad & V(\pcgft)=V(\pft) \,. \label{eq:PCS-1}
\end{aligned}
\end{align}
Under the optimal intervention, $\pcs_{X_{0..\ft},Y_{0..\ft}}$, all meaningless bits of transfer entropy are scrambled
while all remaining transfer entropy is meaningful. We use this to define  %
the \textbf{amount of observed semantic information} as the amount of transfer entropy
under the optimal intervention,
\begin{align}
\SO & =\sum_{t=0}^{\ft-1}\TE[\pcs] \,.
\label{eq:SSopt-1}
\end{align}
Finally,
we define 
the \textbf{semantic efficiency} of observed semantic
information as the ratio of the amount of observed semantic information to the overall transfer entropy,
\[
\pO:=\frac{\SO}{\sum_{t=0}^{\ft-1}\TE[p]}\in[0,1]
\]
Semantic efficiency quantifies %
which portion of transfer entropy determines the system's viability at time $\ft$.
It is non-negative due the non-negativity of transfer entropy.  It is upper bounded by 1 because 
the actual distribution over system-environment trajectories, $p_{X_{0..\ft},Y_{0..\ft}}$, is part of the domain of the minimization in \cref{eq:PCS-1} (corresponding to any $\cg$ which is a one-to-one function), thus
the amount observed semantic information $\SO$ will always be less
than the actual amount of transfer entropy $\TE[p]$.

We now use the fact that $\pcs$ contains only meaningful bits of transfer entropy to define  both the semantic content and pointwise measures of observed semantic information.
Note that 
transfer entropy at time $t$ can be written as
\begin{multline*}
\TE[\pcs] =\\
\sum_{x_{t},y_{t}, x_{t+1}} \pcs(x_{t},y_{t}, x_{t+1}) \log\frac{\pcs(y_{t}\vert x_{t},x_{t+1})}{\pcs(y_{t}\vert x_{t})} \,.
\end{multline*}
We define the \textbf{semantic content of the transition} $x_{t}\mapsto x_{t+1}$ as
the conditional distribution $\pcs(y_{t}\vert x_{t},x_{t+1})$ for all $y_t \in Y$. {This conditional distribution 
captures only those correlations between $(x_t,x_{t+1})$ and $Y_t$ 
that contribute to the system's viability.}
Similarly, we define \textbf{pointwise observed semantic information} using ``pointwise'' 
measures of transfer entropy \cite{lizier2008local,lizier2014measuring} under $\pcs$. In particular, the pointwise observed semantic information for the transition $x_t \mapsto x_{t+1}$ can be defined as
\[
\sO(y_t\vert x_t,x_{t+1}) := \log\frac{\pcs(y_{t}\vert x_{t},x_{t+1})}{\pcs(y_{t}\vert x_{t})}.
\]


It is of interest to define the \textbf{thermodynamic multiplier} for observed semantic information, so as to  compare the viability value of transfer entropy to the cost of acquiring that transfer entropy.  However, there are different ways of quantifying the thermodynamic cost of acquiring transfer entropy, which depend on the particular way that the measurement process is operationalized~\cite{prokopenko2013thermodynamic,horowitz_second-law-like_2014,cafaro_thermodynamic_2016,spinney_transfer_2016,spinney_transfer_2017}.  Because this thermodynamic analysis is more involved than the one for stored semantic information, we leave it for future work.

\subsection{Other kinds of semantic information\label{subsec:Other-kinds-of}}

We have discussed semantic information
defined relative to two measures of syntactic information: mutual
information at $t=0$, and transfer entropy incurred over the course
of $t\in[0..\ft]$. In future work, a similar approach can be
used to define the semantic information relative to other measures of syntactic
information.  For example, one could consider the semantic information
in the transfer entropy from the system to the environment, which would reflect how much ``observations by the environment'' affect the viability
of the system (an example of a system with this kind of semantic information is a human coupled to a so-called ``artificial pancreas''
\cite{doyle2014closed}, a medical device which measures a person's blood
glucose and automatically delivers necessary levels of insulin).
Alternatively,
one might evaluate how mutual information (or transfer entropy, etc.) between internal subsystems of system \sysX affect
the viability of the system. This would uncover ``internal'' semantic
information which would be involved in internal self-maintenance processes, such as homeostasis.


\section{Automatic identification of initial distributions, timescales, and decompositions of interest\label{subsec:Maximization}}

Our measures of semantic information depend on: 1) the decomposition of the world into the system \sysX and the environment \sysY; 2) the timescale $\ft$; and 3) the initial distribution over joint states of the system and environment. 
The factors generally represent ``subjective''
choices of the scientist, indicating  for which systems, temporal scales, and initial conditions the scientist wishes to quantify semantic information.

However, it is also possible to select these factors in a more ``objective''
manner, in particular by choosing decompositions, timescales, and initial distributions for which  semantic information measures --- such as the value of information or the amount of semantic information --- are maximized.

For example, consider fixing a particular timescale $\ft$ and a particular decomposition into system/environment, and then identifying the  initial distribution
 which maximizes the viability value of stored semantic information,
\begin{align}
\p^\star_{X_0,Y_0} \in \argmax_{q_{X_0,Y_0}} \; \VtotS(q_{X_0,Y_0}) \,,
\label{eq:eqps}
\end{align}
where we have made the dependence of $\Vtot$ on the initial distribution explicit in \cref{eq:eqps}, but left implicit its dependence on the timescale $\ft$ and the decomposition into \sysX and \sysY.
Given the intrinsic dynamics of the system and environment, $\p^\star_{X_0,Y_0}$ captures the initial distribution that the system  is ``best fit for'' in an informational sense, i.e., the distribution under which the system most benefits from having syntactic information about the environment. One can then define various other semantic information measures, such as the amount of semantic information and the semantic
content of particular states, relative to $\p^\star_{X_0,Y_0}$, rather than some exogenously specified initial distribution. For instance, the semantic content of some system state $x\in X$ under $\p^\star_{X_0,Y_0}$ represents the conditional distribution over environmental states that, given the dynamics of system and environment, $x$ is ``best fit to represent'' in terms of maximizing viability value.

One can also maximize the value of information (or other measures) over timescales $\ft$ and system/environment decompositions of the world, so as  to automatically detect subsystems and temporal
scales that exhibit large amounts of semantic information.
As mentioned in the Introduction,  our work is conceptually inspired by work on autonomous agents, and our approach in fact suggests a possible formal and quantitative definition
of \textbf{autonomous agency}:  a physical system is an autonomous agent to the extent that it has a large measure of semantic information.  From this point of view, finding timescales and system/environment decompositions that maximize measures of semantic information provides a way to automatically identify agents in the physical world 
(see also~\cite{polani2016towards,biehl2017action,balduzzi2011detecting,krakauer_information_2014}).
Exploring these possibilities, including which semantic information measures (value of information, the amount of semantic information, thermodynamic multiplier, etc.) are best for automatically identifying agents, remains for future work.

%
%
%
%
%
%


\section{Conclusion and Discussion\label{sec:discussion}}

In this paper, we propose a definition of semantic information as the syntactic
information between a physical system and its environment that is causally necessary for maintaining the system's existence.  We consider two particular measures of semantic information: stored semantic information, which is based on the mutual information between system and environment at $t=0$, and observed semantic information, which is based on the transfer entropy exchanged between system and environment over $t\in[0,\ft]$.  



%
%
%
%

Our measures possess several features that have been proposed as desirable characteristics of any measure of semantic information  in the philosophical literature
\cite{godfrey-smith_information_2007,sterelny_sex_1999,godfrey-smith_biological_2016}. Unlike syntactic information, semantic information should be able to be ``mistaken'', i.e., to  ``misrepresent'' the world. This emerges naturally in our framework whenever information has a negative viability value (i.e., when the system uses information in a way that actually hurts its ability to maintain its own existence). Furthermore, a notion of semantic information between a system and environment should be fundamentally \emph{asymmetrical} (unlike some measures of syntactic information, such as mutual information). For instance, a chemotactic bacterium
swimming around a nutrient solution is presumed to have semantic
information about its environment, but the environment is not expected to have semantic information about the bacteria.
Our measures of semantic information are fundamentally asymmetrical --- even when defined relative to a symmetric syntactic information measure like mutual information --- because they are defined in terms of their contribution to the viability of the system, rather than the environment.

Our framework does not require the system of interest to be decomposed into separate degrees of freedom representing  ``sensors'' vs. ``effectors'' (or ``membrane'' vs. ``interior'',
``body'' vs. ``brain'', etc.). This is advantageous because such distinctions may be
difficult or  impossible to define for certain systems. Our framework also side-steps questions of what type  of ``internal models'' or ``internal representations'', if any, are employed by the system. Instead, our definitions of semantic information, including the semantic content of particular states of the system, are grounded  in the intrinsic dynamics of the system and environment.

As mentioned, we do not assume that the system of interest is an organism.  At the same time, 
in cases where the system is in fact an organism (or an entire population of organisms) undergoing an evolutionary process, there are promising connections between our approach and information-theoretic ideas in theoretical biology.  For instance, various ways of formalizing fitness-relevant information in biology~\cite{adami2004information,donaldson2010fitness,krakauer_information_2014} appear conceptually, and perhaps formally, related to our definitions of semantic information. Exploring  such connections remains for future work.

Organisms are, of course, {the} prototypical self-maintaining systems, and will generally have high levels of both stored and observed semantic information.   This suggests that our measures of semantic information may be useful as part of quantitative, formal definitions of life.  In particular, we suggest that having high levels of semantic information is a necessary, though perhaps not sufficient, condition for any physical system to be alive.

%

%
%
%

%
%
%

%
%

%

 %

%

%
%
%
%
%
%
%
%
%
%
%
%
%
%
%
%
%
%
%
%
%
%
%
%
%
%
%
%
%
%
%
%
%

%

%

%
%
%
%
%
%
%
%
%
%
%
%
%
%
%


\ifarxiv
\begin{acknowledgments}
\else
\subsubsection*{Acknowledgments}
\fi
We would like to thank Carlo Rovelli, Daniel Polani,
Jacopo Grilli, Chris Kempes, and Mike Price, along with Luis Bettencourt, Daniel
Dennett, Peter Godrey-Smith, Chris Wood, and other participants in
the Santa Fe Institute's ``Meaning of Information'' working group
for helpful conversations. We would also like to thank the anonymous reviewers for helpful suggestions. We thank the Santa Fe
Institute for helping to support this research. This paper was made
possible through Grant No. FQXi-RFP-1622 from the FQXi foundation,
and Grant No. CHE-1648973 from the U.S. National Science Foundation.
\ifarxiv
\end{acknowledgments}
\fi

\ifarxiv
\bibliographystyle{ieeetr}
\else
\bibliographystyle{unsrtnat}
\fi
\bibliography{meaning}

\appendix


\renewcommand\thefigure{\thesection.\arabic{figure}}

\section{Relationship between entropy and probability of being in viability set}
\label{app:entbound}

\newcommand{\shortin}{\!\!\in\!}
\newcommand{\shortnotin}{\!\!\not\in\!}

Imagine that $\Aset\subseteq X$ is some set of desirable states, which we call the \emph{viability set}.  Assume that
$\left|\Aset\right|\ll\left|X\right|$. Here we show that entropy bounds the probability that \sysX is in set $\Aset$ as
\begin{equation}
p(X \shortin \Aset)=\sum_{x\in\Aset}p(x)\lesssim\frac{\log\left|X\right|-S(p(X))}{\log\left|X\right|-\log\left|\Aset\right|}\label{eq:ent-bound}
\end{equation}

To demonstrate this, let $\mathbf{1}_\Aset(x)$ be the indicator function for set $\Aset$, so that $\mathbf{1}_\Aset(x)$ is equal to 1 when $x\in \Aset$, and 0 otherwise.  Using the chain rule for entropy, we write
\begin{align}
S(p(X)) & =S(p(X,\mathbf{1}_\Aset(X)))\nonumber \\
 & =S(p(X\vert \mathbf{1}_\Aset(X)))+S(\mathbf{1}_\Aset(X))\nonumber \\
 & \le S(p(X\vert \mathbf{1}_\Aset(X)))+1\label{eq:chainruleapp}
\end{align}
In the last line, we use the fact that the maximum entropy of a binary
random variable, such as $\mathbf{1}_\Aset(X)$, is 1 bit.

We now rewrite the conditional entropy as
\begin{align}
& S(p(X\vert \mathbf{1}_\Aset(X))) \nonumber \\
& = p(X \shortin \Aset) S(X\vert X \shortin \Aset)+ (1-p(X \shortin \Aset)) S(X\vert X \shortnotin \Aset)\nonumber \\
 & \le p(X \shortin \Aset)\log\left|\mathcal{A}\right|+(1-p(X \shortin \Aset))\log\left|X\backslash\mathcal{A}\right|\label{eq:uppboundapp}
\end{align}
where we've used the fact that entropy of any distribution over a
set of size $n$ is upper bounded by $\log n$ (as achieved by the
uniform distribution over that set). Combining with \cref{eq:chainruleapp}
gives
\begin{align*}
S(p(X)) \! \le \! p(X \shortin \Aset)\left(\log\left|\mathcal{A}\right| \!-\! \log\left|X\backslash\mathcal{A}\right|\right)+
\log\left|X\backslash\mathcal{A}\right|+1
\end{align*}

Rearranging gives
\begin{align*}
p(X \in \Aset) & \le\frac{-S(p(X))+\log\left|X\backslash\mathcal{A}\right|+1}{\log\left|X\backslash\mathcal{A}\right|-\log\left|\Aset\right|}\\
 & =1-\frac{S(p(X))-\log\left|\Aset\right|-1}{\log\left|X\backslash\mathcal{A}\right|-\log\left|\Aset\right|}\\
 & \le1-\frac{S(p(X))-\log\left|\Aset\right|-1}{\log\left|X\right|-\log\left|\Aset\right|}\\
 & \approx1-\frac{S(p(X))-\log\left|\Aset\right|}{\log\left|X\right|-\log\left|\Aset\right|}\\
 & =\frac{\log\left|X\right|-S(p(X))}{\log\left|X\right|-\log\left|\Aset\right|}
\end{align*}
where we've dropped the $\frac{1}{\log\left|X\backslash\Aset\right|-\log\left|\Aset\right|}$
term.

Thus, as entropy goes up, the probability concentrated within any
small set goes down. 

\clearpage

\onecolumngrid
\section{Model of food-seeking agent\label{app:Food-seeking-Agent-Model}}

\begin{figure*}
\includegraphics[width=5in]{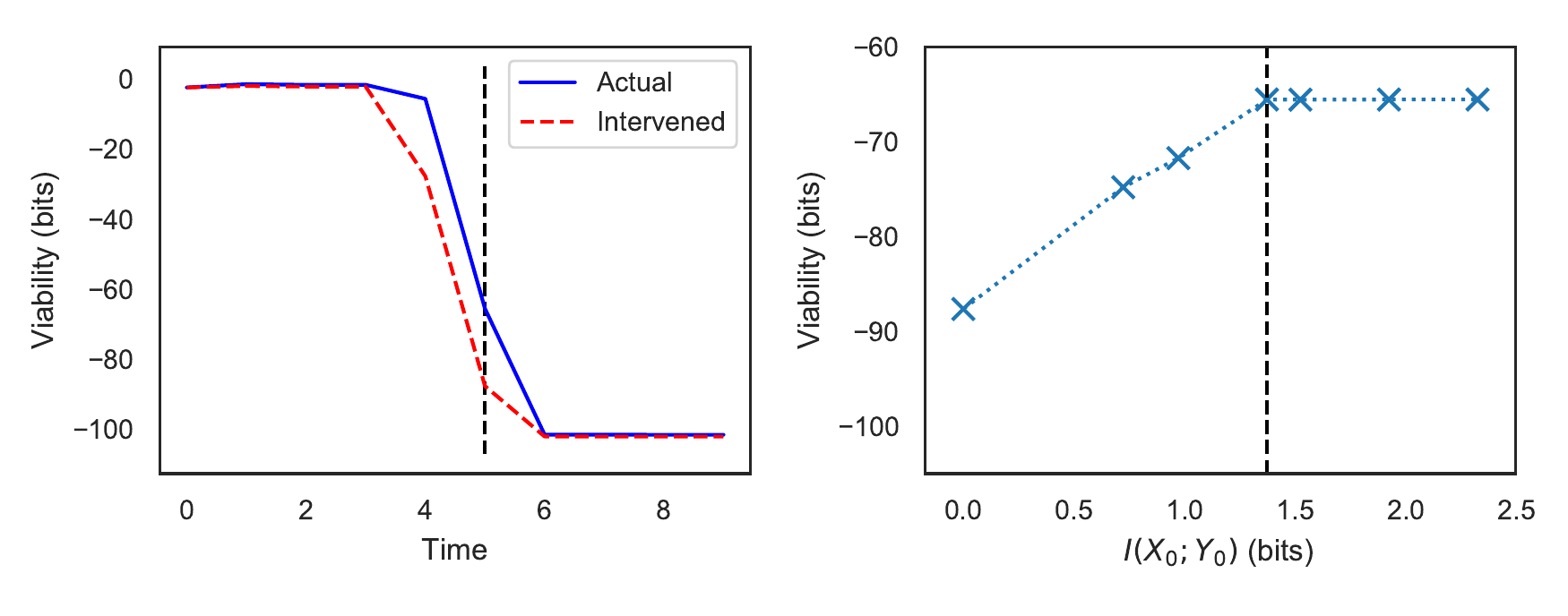}

\caption{Illustration of our approach using a simple model of a food-seeking
system. Under the actual distribution, the
system has perfect knowledge of the location of food at $t=0$. On
the left, we plot viability values over time under both the actual
and (fully scrambled) intervened distributions. The vertical dashed line corresponds to our timescale
of interest ($\protect\ft=5$ timesteps). On the right, we plot the
information/viability curve for $\protect\ft=5$ ($\times$'s are actual
points on the curve, dashed line is interpolation). 
The vertical dashed line indicates the amount of stored semantic information. See text for details.
\label{fig:app-stored-agent-figA}}
\end{figure*}

\begin{figure*}
\includegraphics[width=5in]{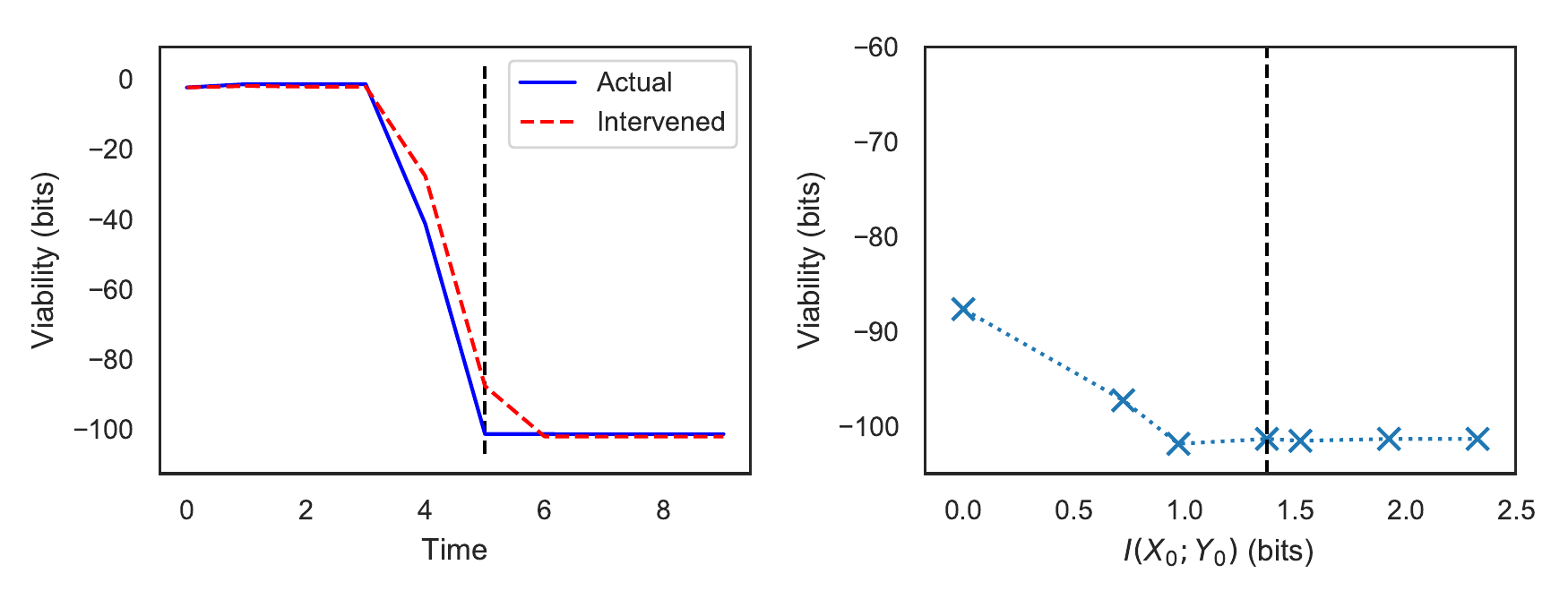}

\caption{Illustration of our measure with a simple model of a system which moves \emph{away} from where it believes food to be located.  On
the left, we plot viability values over time under both the actual
and (fully scrambled) intervened distributions. The vertical dashed line corresponds to our timescale
of interest ($\protect\ft=5$ timesteps). On the right, we plot the
information/viability curve for $\protect\ft=5$ ($\times$'s are actual
points on the curve, dashed line is interpolation). 
The vertical dashed line indicates the amount of stored semantic information. See text for details.
\label{fig:app-stored-agent-figB}}
\end{figure*}

\twocolumngrid

In this section, we describe our model of a simple food-seeking
system.

In this model, the state space of the environment \sysY consists of $Y = \left\{ 1..n\right\} \cup\left\{ \varnothing\right\} $,
representing the location of a single unit of food along 1 spatial dimension,
or the possible lack of food ($\varnothing$). The state space of the agent (i.e., the system \sysX) consists
of three separate degrees of freedom, indicated as $X= X^{loc}\times X^{level}\times X^{target}$.
$X^{loc}=\left\{ 1..n\right\}$ represents the spatial location
of the agent out of $n$ possible locations. $X^{level} = \{0..l_{max}\}$
represents the ``satiation level'' of the agent, ranging from ``fully fed'' ($l_{max}$) to ``dead'' (0). $X^{target} = \{1..n\}\cup\left\{\varnothing\right\}$
represents the agent's internal information about the location of
food in the environment ($\varnothing$ corresponding to
information that there is no food).

The dynamics are such that, as long as the agent is not ``dead''
($X^{level}\ne0$), the agent moves toward $X^{target}$.  If the agent
reaches a location sufficiently close to the food ($\left|X^{loc}-Y\right|\le1$),
the agent ``eats the food'', meaning that satiation level of the
agent is changed to $l_{max}$. Otherwise, the satiation level drops
by one during every timestep. 
The food stays in the same place unless it gets eaten, or unless it spontaneously degrades (goes to $\varnothing$) which happens with a small probability in each step.  
The agent never changes its target belief.
All states are assigned free energy values, for which the dynamics obey local detailed balance.

Initially, the agent is located
at the center spatial location ($X_{0}^{loc}=\left\lfloor n/2\right\rfloor $),
the satiation level is maximal $X_0^{level}=l_{max}$, 
the food location is uniformly distributed over $1..n$, and
the agent has perfect information about the location of the food, 
$p(x_{0}^{target}\vert y_{0})=\delta(x_{0}^{target},y_{0})$.

We assume that the
state space of the agent corresponds to a set coarse-grained macrostates. Formally, we write this as 
$X = f(Z)$, where $Z$ is a random variable indicating the microstate of \sysX and $f$ is a function that maps from microstates to macrostates.  The entropy of any microstate distribution $p_{Z_\ft}$ can be written as 
\begin{align*}
S(p_{Z_\ft}) & = S(p_{Z_\ft,X_\ft}) \\
& = S(p_{X_\ft}) + S(p_{Z_\ft|X_\ft}) \\
& = S(p_{X_\ft}) + \sum_{x_\ft} p(x_\ft) S(p_{Z_\ft|f(Z_\ft) = x_{\ft}}) \,,
\end{align*}
We assume that within each macrostate, the microstate distribution relaxes instantly to some local equilibrium, so that each ``internal entropy'' term $S(p_{Z_\ft|f(Z_\ft) = x_{\ft}})$ is  constant, which we indicate as $S_\text{int}(x_\ft)$. Combining, we compute our negative entropy viability function as
\[
V(\pft) = S(\pft) + \sum_{x_\ft} p(x_\ft) S_\text{int}(x_\ft) \,.
\]
In this particular model, we take the
internal entropy of all macrostates to be 0, except for any macrostate which has $X^{level}=0$
(i.e., the agent is ``dead''), in which case the internal entropy
is $S_\text{dead}$ bits. Essentially, this means that the system equilibrates instantly
within the dead macrostate, and that the dead macrostate has a large
internal entropy (i.e., there are many more ways of being dead than
not). 

To avoid having results that are sensitive to numerical errors, we ``smooth'' the information/viability curve by rounding all viability and mutual information values to 5 decimal places.

\cref{fig:app-stored-agent-figA} shows the results for parameters $n=5$, $l_{max}=5$, $S_\text{dead}=100$ bits, 
and timescale $\ft=5$. 
The total amount of mutual information is $\log_2 5 \approx 2.32$ bits, while the total amount of semantic information is only $\approx 1.37$ bits, which gives a semantic efficiency value of $\nS \approx 0.6$.  This occurs because if the food is initially in locations $\{2,3,4\}$, the agent is
close enough to eat it immediately, and knowing in which of these
3 locations the food is located does not affect viability.
The viability value of information is $\VtotS \approx 22.1$ bits, giving a thermodynamic multiplier of $\nS \approx9.5$. The model is also discussed in \cref{sec:stored-example} in the main text.

We also analyze a different model, in which the agent's dynamics are such that it moves \emph{away} from the target in each timestep, until it reaches the edges of world ($X^{loc}=1$ or $X^{loc}=n$) and stays there.  The agent still dies if it does not eat food for some number of timesteps.  As before, the agent begins initially with perfect information about the location of the food.  In this case, information about the world actually hurts the agent's ability to maintain its own existence, leading to a negative viability value of information.

\cref{fig:app-stored-agent-figB} shows the results for this model, using the same parameter values as before ($n=5$, $l_{max}=5$, $S_\text{dead}=100$ bits, 
and timescale $\ft=5$). 
The total amount of mutual information is again $\log_2 5 \approx 2.32$ bits, and the total amount of semantic information is again $\approx 1.37$ bit (if the food is initially in locations $\{2,3,4\}$, the system is
close enough to eat it immediately, and knowing in which of these
3 locations the food is located does not affect viability).
This gives a semantic efficiency value of $\nS \approx 0.6$
Unlike the food-seeking agent, the viability value of information in this case is $\VtotS \approx -13.7$ bits, giving a thermodynamic multiplier of $\nS \approx -5.9$.

A Python implementation of these models is available at \url{https://github.com/artemyk/semantic_information/}.

\clearpage
\end{document}